\documentclass[twocol]{ametsoc}

\journal{mwr}
\usepackage{float}
\usepackage{amsmath}
\usepackage{amsfonts}
\usepackage{nicefrac}
\usepackage{amssymb}
\usepackage{color}
\usepackage{cases}
\usepackage{pifont}
\usepackage{array}
\usepackage{ulem}
\usepackage{todonotes}
\usepackage[boxed]{algorithm2e}
\renewcommand{\v}[1]{\ensuremath{\mathbf{#1}}}
\newcommand{\gv}[1]{\bm{#1}}

\newcommand{\black}[1]{\textcolor{black}{#1}}

\newcolumntype{C}[1]{>{\centering\let\newline\\\arraybackslash\hspace{0pt}}m{#1}}

\usepackage{tabularx}
\DeclareMathOperator*{\argmin}{arg\,min}
\newcommand{\transp}{\mathrm{T}}
\newcommand{\norm}[1]{\left\lVert#1\right\rVert}

\bibpunct{(}{)}{;}{a}{}{,}

\title{A Review of Innovation-Based Methods to Jointly Estimate Model and Observation Error Covariance Matrices in Ensemble Data Assimilation}

\authors{Pierre Tandeo\correspondingauthor{Dept. Signal \& Communications, IMT Atlantique, 655 Avenue du Technop\^ole, 29200 Plouzan\'e, France}}

\affiliation{IMT Atlantique, Lab-STICC, UMR CNRS 6285, F-29238, France \& RIKEN Center for Computational Science, Kobe, Japan}
\email{pierre.tandeo@imt-atlantique.fr}

\extraauthor{Pierre Ailliot}
\extraaffil{LMBA, UMR CNRS 6205, University of Brest, France}

\extraauthor{Marc Bocquet}
\extraaffil{CEREA Joint Laboratory \'Ecole des Ponts ParisTech and EDF R\&D, Universit\'e Paris-Est, Champs-sur-Marne, France}

\extraauthor{Alberto Carrassi}
\extraaffil{Dept. of Meteorology and National Centre for Earth Observation, University of Reading, UK
\& Mathematical Institute, University of Utrecht, Netherlands}

\extraauthor{Takemasa Miyoshi}
\extraaffil{RIKEN Center for Computational Science, Kobe, Japan}

\extraauthor{Manuel Pulido}
\extraaffil{Universidad Nacional del Nordeste and CONICET, Corrientes, Argentina \& Department of Meteorology, University of Reading, UK}

\extraauthor{Yicun Zhen}
\extraaffil{IMT Atlantique, Lab-STICC, UMR CNRS 6285, F-29238, France}

\abstract{Data assimilation combines forecasts from a numerical model with observations. Most of the current data assimilation algorithms \black{consider} the model and observation error terms as additive Gaussian noise, \black{specified by their} covariance matrices Q and R\black{, respectively}. \black{These error covariances, and specifically their respective amplitudes, determine the weights given to the background (i.e., the model forecasts) and to the observations in the solution of data assimilation algorithms (i.e., the analysis).} Consequently, Q and R matrices \black{significantly impact the accuracy of the analysis.} This review aims to present \black{and to discuss, with a unified framework,} different methods to jointly estimate \black{the} Q and R matrices using ensemble-based data assimilation techniques. Most of the methodologies developed to date use the innovations, defined as differences between the observations and \black{the projection of the forecasts onto the observation space. These methodologies are based on two main statistical criteria}: (i) the method of moments, in which the theoretical and empirical moments of the innovations are assumed to be equal, and (ii) methods \black{that use the likelihood of the observations, themselves contained in the innovations. The reviewed methods assume that innovations are Gaussian random variables, although extension to other distributions is possible for likelihood-based methods. The methods also show some differences in terms of levels of complexity and applicability to high-dimensional systems. The conclusion of the review discusses the key challenges to further develop estimation methods for Q and R. These challenges include taking into account time-varying error covariances, using limited observational coverage, estimating additional deterministic error terms, or accounting for correlated noises.}}

\begin{document}

\maketitle

%%%%%%%%%%%%%%%%%%%%%%%%%%%%%%%%%%%%%%%%%%%%%%%%%%%%%%%%%%%%%%%%%%%%%
% INTRODUCTION
%%%%%%%%%%%%%%%%%%%%%%%%%%%%%%%%%%%%%%%%%%%%%%%%%%%%%%%%%%%%%%%%%%%%%

\section{Introduction}

In meteorology and other environmental sciences, an important challenge is to estimate the state of the system as accurately as possible. \black{In} meteorology, this state includes pressure, humidity, temperature and wind at different locations and elevations \black{in the atmosphere}. Data assimilation (hereinafter DA) refers to mathematical methods that use both model predictions (also called background information) and partial observations to retrieve the current state vector with its associated error. An accurate estimate of the current state is crucial to get good forecasts, and it is particularly so whenever the system dynamics is chaotic, such as it is the case for the atmosphere.

The performance of a DA system to estimate the state depends on the accuracy of the model predictions, the observations, and their associated error terms. A simple, popular and mathematically justifiable way of modeling these errors is to assume them to be independent and unbiased Gaussian white noise, with covariance matrices $\v{Q}$ for the model and $\v{R}$ for the observations. \black{Given the aforementioned importance of $\v{Q}$ and $\v{R}$ in estimating the analysis state and error, a number of studies dealing with this problem has arisen in the last decades.} This review work presents and summarizes the different techniques used to estimate simultaneously the $\v{Q}$ and $\v{R}$ covariances. Before discussing the methods to achieve this goal, the mathematical formulation of DA is briefly introduced.

\subsection{Problem statement}\label{subsec_problem_notations} % , notations and role of Q and R matrices

\begin{figure}[!ht]
  \centerline{\includegraphics[width=0.5\textwidth]{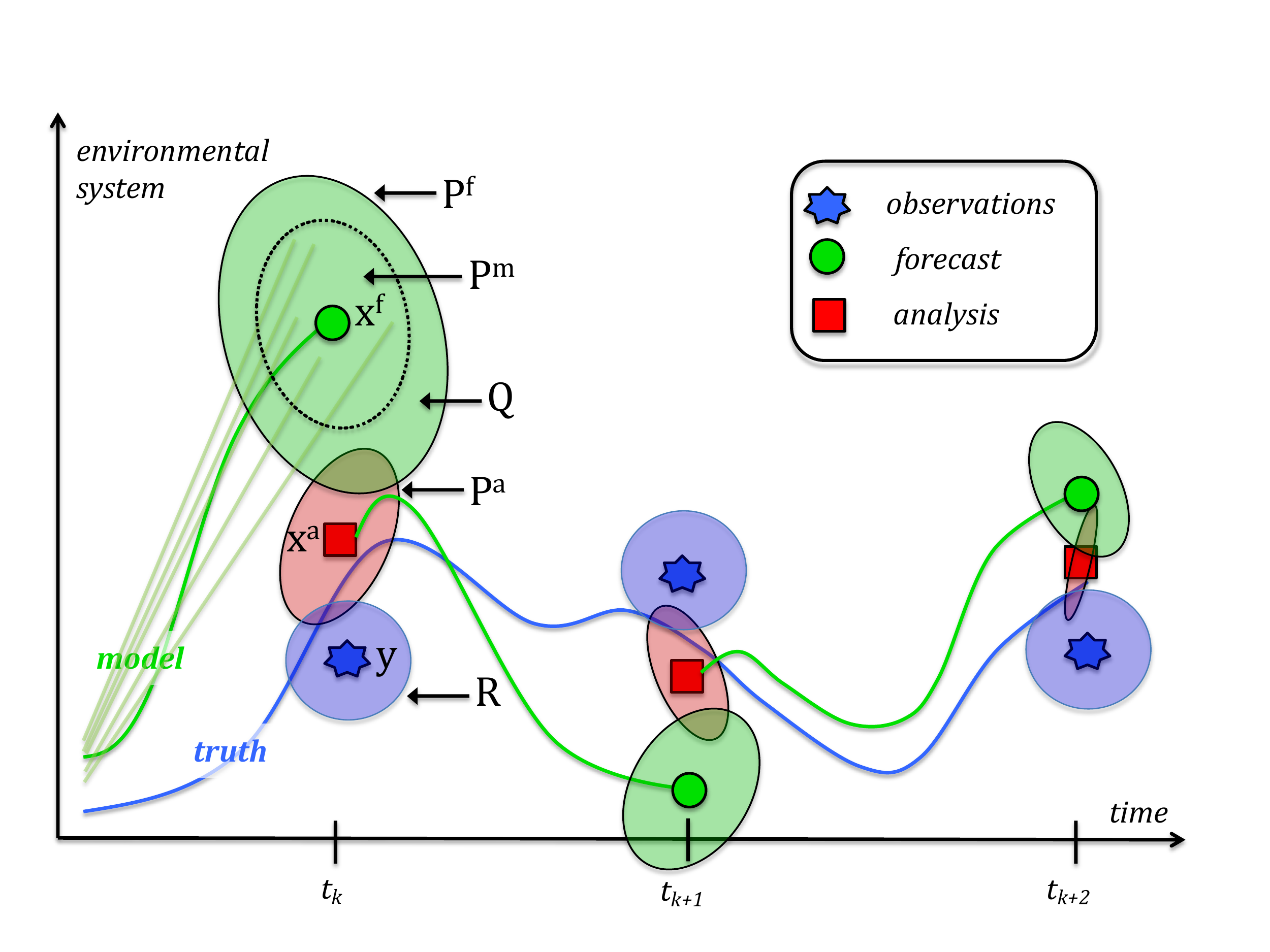}}\caption{Sketch of sequential and ensemble data assimilation algorithms in the observation space (i.e., in the space of the observations $\v{y}$), where the observation operator $\mathcal{H}$ is omitted for simplicity. The ellipses represent the forecast $\v{P}^f$ and analysis $\v{P}^a$ error covariances, while the model $\v{Q}$ and observation $\v{R}$ error covariances are the unknown entries of the state-space model in \black{Eqs.~(\ref{eq_state}) and (\ref{eq_obs})}. The forecast error covariance matrix is written $\v{P}^f$ and is the sum of $\v{P}^m$, the forecasted state $\v{x}^f$ spread, and the model error $\v{Q}$. This scheme is a modified version based on Fig.~1 from \cite{Carrassi2018DataPerspectives}.}\label{fig_schema_kalman_Pf_Q_R}
\end{figure}

Hereinafter, the unified DA notation proposed in \cite{Ide1997UnifiedVariational} \black{is} used\footnote{Other notations are also used in practice}. DA algorithms are used to estimate the state of a system, $\v{x}$, conditionally on observations, $\v{y}$. A classic strategy is to use sequential and ensemble DA frameworks, as illustrated in Fig.~\ref{fig_schema_kalman_Pf_Q_R}, and to combine two sources of information: model forecasts (in green) and observations (in blue). The ensemble framework uses different realizations, also called members, to track the state of the system at each assimilation time step.

The forecasts of the state are based on the usually incomplete and approximate knowledge of the system dynamics. The evolution of the state from time $k-1$ to $k$ is given by the model equation:
  \begin{equation}
  \v{x}(k) = \mathcal{M}_k \left(\v{x}(k-1) \right) + \gv{\eta}(k), \label{eq_state}
\end{equation}
where the model error $\gv{\eta}$ implies that the \black{dynamic model operator $\mathcal{M}_k$ is} not perfectly known. Model error is usually assumed to follow a Gaussian distribution with zero mean (i.e., the model is unbiased) and covariance $\v{Q}$. The dynamic model operator $\mathcal{M}_k$ in Eq.~(\ref{eq_state}) has also an explicit dependence on $k$, because it may depend on time-dependent external forcing terms. At time $k$, the forecasted state is characterized by the mean of the forecasted states, $\v{x}^f$, and its uncertainty matrix, namely $\v{P}^f$, which is also called the background error covariance matrix\black{, and noted $\v{B}$ in DA}.

The forecast covariance $\v{P}^f$ is determined by two processes. The first is the uncertainty propagated from $k-1$ to $k$ by the model $\mathcal{M}_k$ (the green shade within the dashed ellipse in Fig.~\ref{fig_schema_kalman_Pf_Q_R}, and denoted by $\v{P}^m$). The second process is the model error covariance $\v{Q}$ accounted by the noise term at time $k$ in Eq.~(\ref{eq_state}). Given that model error is largely unknown and originated by various and diverse sources, the matrix $\v{Q}$ is also poorly known. Model error sources encompass the model $\mathcal{M}$ deficiencies to represent the underlying physics, including deficiencies in the numerical schemes, the cumulative effects of errors in the parameters, and the lack of knowledge of the unresolved scales. Its estimation is a challenge in general, but it is particularly so in geosciences because we usually have far fewer observations than those needed to estimate the entries of $\v{Q}$ \citep{Daley1992EstimatingAssimilation,Dee1995On-lineAssimilation}. The sum of the two covariances $\v{P}^m$ and $\v{Q}$ gives the forecast covariance matrix, $\v{P}^f$ (full green ellipse in Fig.~\ref{fig_schema_kalman_Pf_Q_R}). \black{In the illustration given here, a large contribution of the forecast covariance $\v{P}^f$ is due to $\v{Q}$. This situation reflects what is common in ensemble DA, where $\v{P}^m$ can be too small, as a consequence of the ensemble undersampling of the initial condition error (i.e., the covariance estimated at the previous analysis). In that case, inflating $\v{Q}$ could partially compensate for the bad specification of $\v{P}^m$.}

DA uses a second source of information, the observations $\v{y}$, which are assumed to be linked to the true state $\v{x}$ through the time-dependent operator $\mathcal{H}_k$. This step in DA algorithms is formalized by the observation equation:
\begin{equation}
  \v{y}(k) = \mathcal{H}_k \left(\v{x}(k) \right) + \gv{\epsilon}(k), \label{eq_obs}
\end{equation}
where the observation error $\gv{\epsilon}$ describes the discrepancy between what is observed and the truth. In practice, it is important to remove as much as possible the large-scale bias in the observation before DA. Then, it is common to state that the remaining error $\gv{\epsilon}$ follows a Gaussian and unbiased distribution with a covariance $\v{R}$ (the blue ellipse in Fig.~\ref{fig_schema_kalman_Pf_Q_R}). This covariance takes into account errors in the observation operator $\mathcal{H}$, the instrumental noise and the representation error associated with the observation, typically measuring a higher resolution state than the model represents. In practice, a correct estimation of $\v{R}$ \black{that} takes into account all these effects is often challenging \citep{Janjic2018OnAssimilation}.

DA algorithms combine forecasts with observations, based on the model and observation equations, respectively given in Eq.~(\ref{eq_state}) and Eq.~(\ref{eq_obs}). The corresponding system of equations is a nonlinear state-space model. As illustrated in Fig.~\ref{fig_schema_kalman_Pf_Q_R}, this Gaussian DA process produces a posterior Gaussian distribution with mean $\v{x}^a$ and covariance $\v{P}^a$ (red ellipse). The system given in \black{Eqs.~(\ref{eq_state}) and (\ref{eq_obs})} is representative of a broad range of DA problems, as described in seminal papers such as \cite{Ghil1991DataOceanography.pdf}, and still relevant today as referenced by \cite{Houtekamer2016ReviewAssimilation} and \cite{Carrassi2018DataPerspectives}. The assumptions made in \black{Eqs.~(\ref{eq_state}) and (\ref{eq_obs})} about model and observation errors (additive, Gaussian, unbiased, and mutually independent) are strong, yet convenient from the mathematical and computational point of view. Nevertheless, these assumptions are not always realistic in real DA problems. For instance, in operational applications, systematic biases in the model and in the observations are recurring problems. Indeed, biases affect significantly the DA estimations and a specific treatment is required \black{;} see \cite{Dee2005BiasAssimilation} for more details.

From \black{Eqs.~(\ref{eq_state}) and (\ref{eq_obs})}, noting that $\mathcal{M}$, $\mathcal{H}$ and $\v{y}$ are given, the only parameters that influence the estimation of $\v{x}$ are the covariance matrices $\v{Q}$ and $\v{R}$. In practice, these covariances play an important role in DA algorithms. Their importance was early put forward in \cite{hollingsworth1986statistical}, in section~4.1 of \cite{Ghil1991DataOceanography.pdf} and \cite{Daley1991AtmosphericAnalysis} in section~4.9. The results of DA algorithms highly depend on the two error covariance matrices $\v{Q}$ and $\v{R}$, which have to be specified by the users. But in practice, these covariances are not easy to tune. Indeed, their impact is hard to grasp in real DA problems with high-dimensionality and nonlinear dynamics. We thus illustrate the problem with a simple example first.

\begin{figure*}[!ht]
 \centerline{\includegraphics[width=\textwidth]{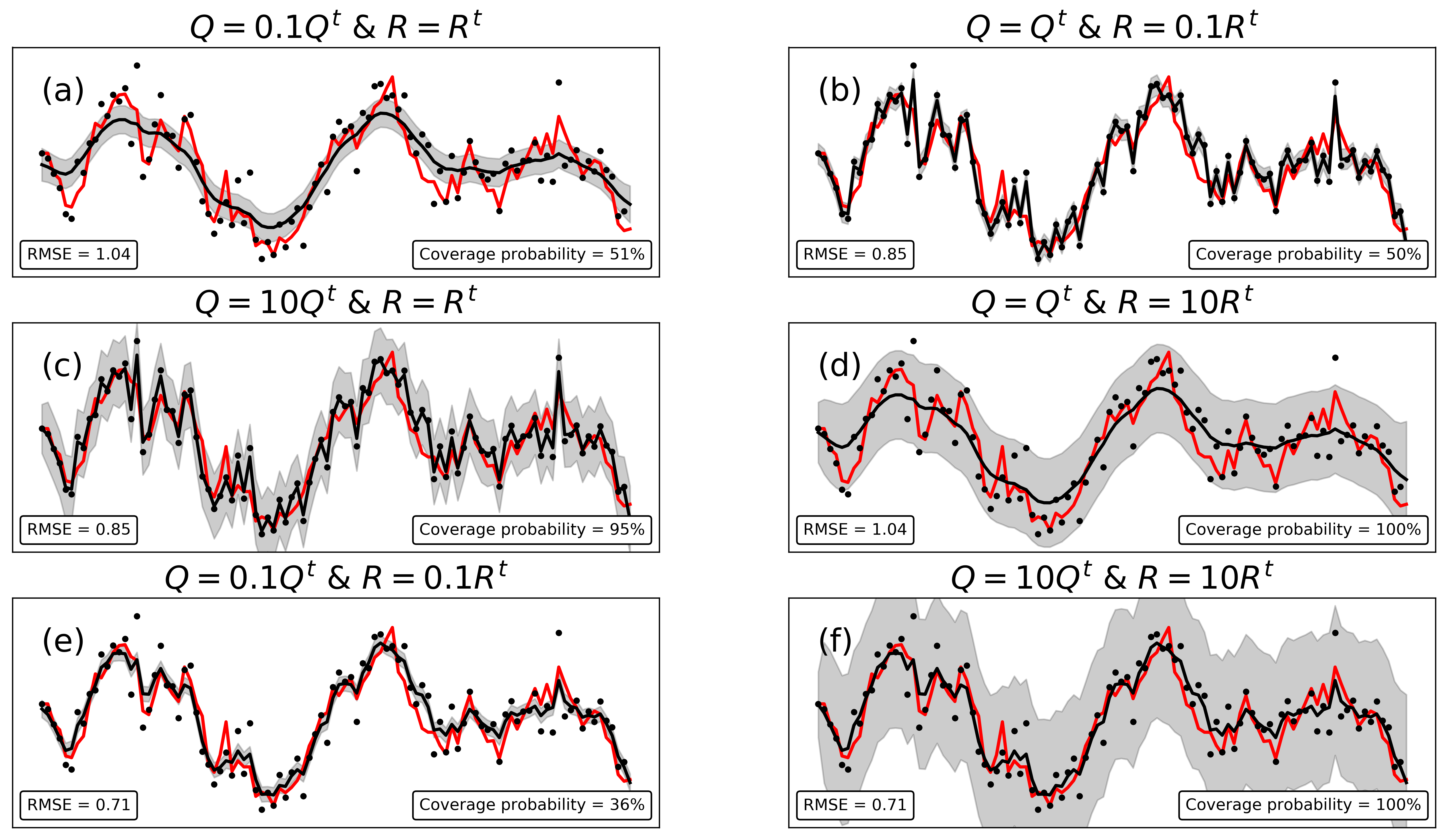}}\caption{Example of a univariate AR(1) process generated using Eq.~(\ref{eq_AR1}) with $Q^t=1$ (red line), noisy observations as in Eq.~(\ref{eq_obs}) with $R^t=1$ (black dots) and reconstructions with a Kalman smoother (black lines and gray $95\%$ confidence interval) with different values of $Q$ and $R$, from $0.1$ to $10$. The optimal values of RMSE and coverage probabilities are, respectively, $0.71$ and $95\%$.}\label{fig_1}
\end{figure*}

\subsection{Illustrative example}\label{subsec_simple_example}

In either variational or ensemble-based DA methods, the quality of the reconstructed state (or hidden) vector $\v{x}$ largely depends on the relative amplitudes between the assumed observation and model errors \citep{Desroziers2001DiagnosisAssimilation}. In \black{Kalman filter based methods}, the signal-to-noise ratio $\norm{\v{P}^f}/\norm{\v{R}}$, where $\v{P}^f$ depends on $\v{Q}$, impacts the \black{Kalman gain}, which gives the relative weights of the observations against the model forecasts. Here, the $\norm{.}$ operator represents a matrix norm. For instance, \cite{Berry2013AdaptiveSystems} used the Frobenius norm to study the effect of this ratio in the reconstruction of the state in toy models.

The importance of $\v{Q}$, $\v{R}$ and $\norm{\v{P}^f}/\norm{\v{R}}$ is illustrated with the aid of a toy example, using a scalar state $x$ and simple linear dynamics. This simplified setup avoids several issues typical of realistic DA applications: the large dimension of the state, the strong nonlinearities and the chaotic behavior. In this example, the dynamic model in Eq.~(\ref{eq_state}) is a \black{first-order} autoregressive model, denoted by AR(1) and defined by
\begin{equation}
x(k) = 0.95 x(k-1) + \eta(k), \label{eq_AR1}
\end{equation}
with $\eta \sim \mathcal{N}\left(0,Q^t\right)$ where the superscript $t$ means ``true'' and $Q^t=1$. Furthermore, observations $y$ of the state are contaminated with an independent additive zero-mean and unit-variance Gaussian noise, such that $R^t=1$ in Eq.~(\ref{eq_obs}) with $\mathcal{H} \left( x \right)=x$. The goal is to reconstruct $x$ from the noisy observations $y$ at each time step. The AR(1) dynamic model defined by Eq.~(\ref{eq_AR1}) has an autoregressive coefficient close to one, representing a process which evolves slowly over time, and a stochastic noise term $\eta$ with variance $Q^t$. Although the knowledge of these two sources of noise is crucial for the estimation problem, in practice identifying them is not an easy task. Given that the dynamic model is linear and the error terms are additive and Gaussian in this simple example, the Kalman smoother provides the best estimation of the state (see section~\ref{sec_EKF_EKS} for more details). To evaluate the effect of badly specified $Q$ and $R$ errors on the reconstructed state with the Kalman smoother, different experiments were conducted with values of $\{0.1, 1, 10\}$ for the ratio $Q/R$ (in this toy example, we use $Q/R$ instead of $\norm{\v{P}^f}/\norm{\v{R}}$ for simplicity).

Figure \ref{fig_1} shows, as a function of time, the true state (red line) and the smoothing Gaussian distributions represented by the $95\%$ confidence intervals (gray shaded) and their means (black lines). We also report the Root Mean Squared Error (RMSE) of the reconstruction and the so-called ``coverage probability'', or percentage of $x$ that falls in the $95\%$ confidence intervals (defined as the mean $\pm 1.96$ the standard deviation in the Gaussian case). In this synthetic experiment, the best RMSE and coverage probability obtained, applying the Kalman smoother with true $Q^t=R^t=1$, are $0.71$ and $95\%$, respectively. Using a small model error variance $Q=0.1 Q^t$ in Fig.~\ref{fig_1}(a), the filter gives a large weight to the forecasts given by the quasi-persistent autoregressive dynamic model. On the other hand, with a small observation error variance $R=0.1 R^t$ in Fig.~\ref{fig_1}(b), excessive weight is given to the observation and the reconstructed state is close to the noisy measurements. These results show the negative impact of independently badly scaled $Q$ and $R$ error variances. In the case of overestimated model error variance as in Fig.~\ref{fig_1}(c), the mean reconstructed state vector and thus its RMSE are identical to Fig.~\ref{fig_1}(b). In the same way, overestimated observation error variance like in Fig.~\ref{fig_1}(d) gives similar mean reconstruction, as in Fig.~\ref{fig_1}(a). These last two results are due to the fact that in both cases, the ratio $Q/R$ are equal, respectively, to $10$ and $0.1$. Now, we consider in Fig.~\ref{fig_1}(e) and Fig.~\ref{fig_1}(f) the case where the $Q/R$ ratio is equal to $1$, but, respectively, using the simultaneous underestimation and overestimation of model and observation errors. In both cases, the mean reconstructed state is equal to that obtained with the true error variances (i.e., RMSE=$0.71$). The main difference is the gray confidence interval, which is supposed to contain $95\%$ of the true trajectory: the spread is clearly underestimated in Fig.~\ref{fig_1}(e) and overestimated in Fig.~\ref{fig_1}(f), with respective coverage probability of $36\%$ and $100\%$.

We used a simple synthetic example, but for large dimensional and highly nonlinear dynamics, such an underestimation or overestimation of uncertainty may have a strong effect and may cause filters to collapse. The main issue in ensemble-based DA is an underdispersive spread, as in Fig.~\ref{fig_1}(e). In that case, the initial condition spread is too narrow, and model forecasts (starting from these conditions) would be similar and potentially out of the range of the observations. In the case of an overdispersive spread, as in Fig.~\ref{fig_1}(f), the risk is that only a small portion of model forecasts would be accurate enough to produce useful information on the true state of the system. This illustrative example shows how important is the joint tuning of model and observation errors in DA. Since the 1990s, a \black{substantial} number of studies have dealt with this topic.

\subsection{Seminal work in the data assimilation community}\label{subsec_seminal_work}

In a seminal paper, \cite{Dee1995On-lineAssimilation} proposed an estimation method for parametric versions of $\v{Q}$ and $\v{R}$ matrices. The method, based on maximizing the likelihood of the observations, yields an estimator which is a function of the innovation defined by $\v{y}-\mathcal{H}(\v{x}^f)$. Maximization is performed at each assimilation step, with the current innovation computed from the available observations. This technique was later extended to estimate the mean of the innovation, which depends on the biases in the forecast and in the observations \citep{Dee1999Maximum-LikelihoodMethodology}. The methodology was then applied to realistic cases in \cite{Dee1999Maximum-LikelihoodApplications}, making the maximization of innovation likelihood a promising technique for the estimation of errors in operational forecasts.

Following a distinct path, \cite{Desroziers2001DiagnosisAssimilation} proposed using the observation-minus-analysis diagnostic. It is defined by $\v{y}-\mathcal{H}(\v{x}^a)$ with $\v{x}^a$ the analysis (i.e., the output of DA algorithms). \black{The authors proposed an iterative optimization technique to estimate a scaling factor for the background $\v{B}=\v{P}^f$ and observation $\v{R}$ matrices}. The procedure was shown to converge to a proper fixed-point. As in Dee's work, the fixed-point method presented in \cite{Desroziers2001DiagnosisAssimilation} is applied at each assimilation step, with the available observations at the current step.

\begin{figure*}[!ht]
\centerline{\includegraphics[width=\textwidth]{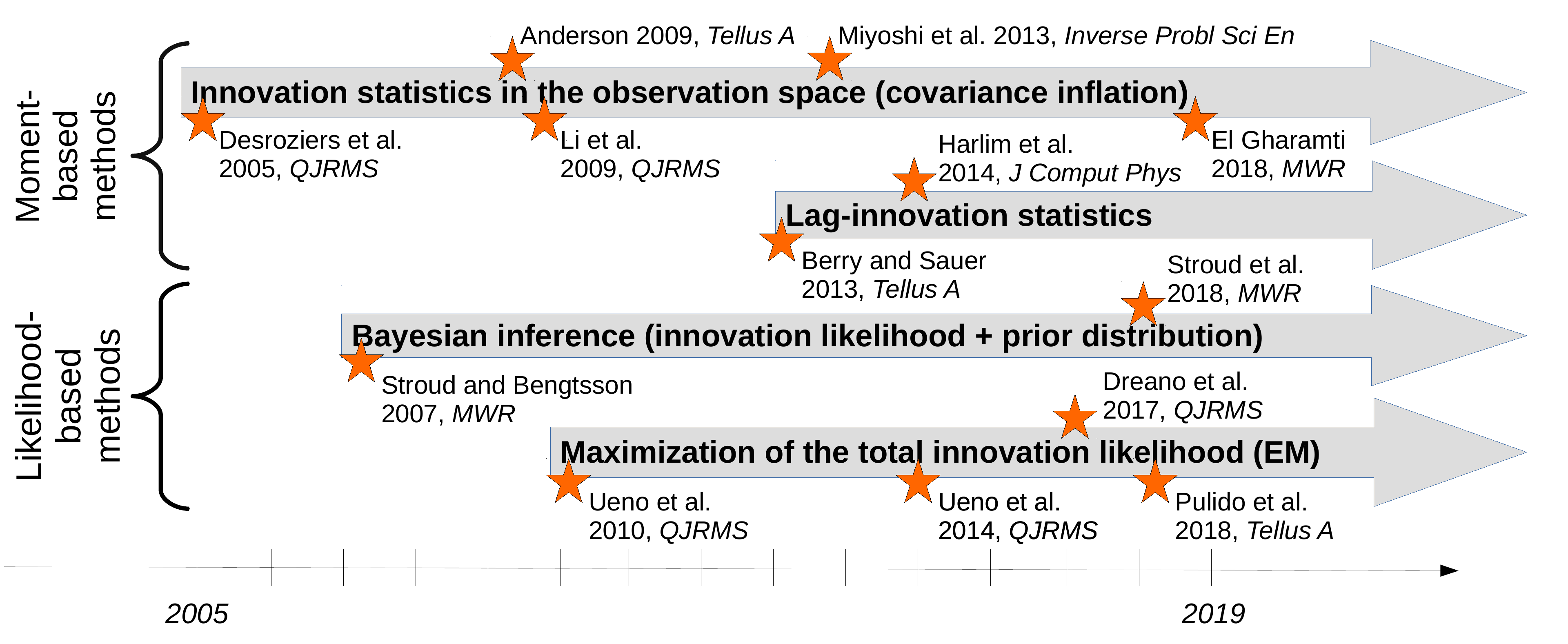}}\caption{Timeline of the main \black{methods} used in geophysical data assimilation for the joint estimation of $\v{Q}$ and $\v{R}$ over the last 15 years. \black{\cite{Dee1995On-lineAssimilation} and \cite{Desroziers2001DiagnosisAssimilation} are not represented here but are certainly the seminal work of this research field in data assimilation.}}\label{fig_timeline}
\end{figure*}

Later, \cite{Chapnik2004PropertiesAssimilation} showed that the maximization of the innovation likelihood proposed by \cite{Dee1995On-lineAssimilation} makes the observation-minus-analysis diagnostic of \cite{Desroziers2001DiagnosisAssimilation} optimal. Moreover, the techniques of \cite{Dee1995On-lineAssimilation} and \cite{Desroziers2001DiagnosisAssimilation} have been further connected to the generalized cross-validation method previously developed by statisticians \citep{wahba1980some}.

\black{These initial studies clearly nurtured the discussion of the estimation of observation $\v{R}$, model $\v{Q}$, or background $\v{B}=\v{P}^f$ error covariance matrices in the modern DA literature. For demonstration purposes, the algorithms proposed in \cite{Dee1995On-lineAssimilation} and \cite{Desroziers2001DiagnosisAssimilation} were tested on realistic DA problems, using a shallow-water model on a plane with a simplified Kalman filter, and using the French ARPEGE three-dimensional variational framework, respectively. In both cases, although good performances have been obtained with a small number of iterations, the proposed algorithms have shown some limits, in particular with regard to the simultaneous estimation of the two sources of errors: observation and model (or background). In this context, \cite{Todling2015AMethod} pointed out that using only the current innovation is not enough to distinguish the impact of $\v{Q}$ and $\v{R}$, which still makes their simultaneous estimation challenging. Given that our preliminary focus here is to review methods for the joint estimate of $\v{Q}$ and $\v{R}$, the work \cite{Dee1995On-lineAssimilation} and \cite{Desroziers2001DiagnosisAssimilation} are not further detailed hereafter. After these two seminal studies, various alternatives were proposed. They are based on the use of several types of innovations and are discussed in this review.}

\subsection{Methods presented in this review}\label{subsec_summary_paper}

The main topic of this review is the \black{``}joint estimation of $\v{Q}$ and $\v{R}$''. Thus, only methods based on this specific goal are presented in detail. A history of what have been, in our opinion, the most relevant contributions and the key milestones for $\v{Q}$ and $\v{R}$ covariance estimation in DA is sketched in Fig.~\ref{fig_timeline}. The highlighted papers are discussed in this review, with a summary of the different methodologies\black{,} given in Table~\ref{tab_summary}. We distinguish four \black{methods} and we can classify them into two categories: those which rely on moment-based methods, and those using likelihood-based methods. Both methods make use of the innovations. The main concepts of the techniques are briefly introduced below.

On the one hand, moment-based methods assume equality between theoretical and empirical statistical moments. A first approach is to study different type of innovations in the observation space (i.e., working in the space of the observations instead of the space of the state). It has been initiated in DA by \cite{rutherford1972data} and \cite{hollingsworth1986statistical}. A second approach extracts information from the correlation between lag innovations, namely innovations between consecutive times. On the other hand, likelihood-based methods aim to maximize likelihood functions with statistical algorithms. One option is to use a Bayesian framework, assuming prior distributions for the parameters of $\v{Q}$ and $\v{R}$ covariance matrices. \black{Another option} is to use the iterative \black{expectation--maximization} algorithm to maximize a likelihood function.

The four methodologies listed in Fig.~\ref{fig_timeline} will be examined in this paper. Before doing that, it is worth \black{mentioning} existing review work that have attempted to summarize the methodologies in DA context and beyond.

\subsection{Other review papers}\label{subsec_other_reviews}

Other review papers on parameter estimation (including $\v{Q}$ and $\v{R}$ matrices) in state-space models have appeared in the statistical and signal processing communities. The first one \citep{Mehra1972ApproachesFiltering} introduces moment- and likelihood-based methods in the linear and Gaussian case \black{(i.e., when $\gv{\eta}$ and $\gv{\epsilon}$ are Gaussians and $\mathcal{M}$ is a linear operator in Eqs.~(\ref{eq_state}) and (\ref{eq_obs}))}. Many extensions to nonlinear state-space models have been proposed since the seminal work of Mehra, and these studies are summarized in the recent review by \cite{Dunik2017NoiseI}, with a focus on moment-based methods and the extended Kalman filter \citep{jazwinski2007stochastic}. The book chapter by \cite{buehner2010error} presents another review of moment-based methods, with a focus on the modeling and estimation of spatial covariance structures $\v{Q}$ and $\v{R}$ in DA with the ensemble Kalman filter algorithm \citep{evensen2009data}.

In the statistical community, the recent development of powerful simulation techniques, known as sequential Monte-Carlo algorithms or particle filters, has led to an extensive literature on the statistical inference in nonlinear state-space models relying on likelihood-based approaches. A recent and detailed presentation of this literature can be found in \cite{Kantas2015OnModels}. However, these methods typically require a large number of particles, which make them impractical for geophysical DA applications.

\black{The review presented here focuses on methods proposed in DA, especially the moment- and likelihood-based techniques which are suitable for geophysical systems (i.e., with high dimensionality and strong nonlinearities).}

\subsection{Structure of this review}

The paper is organized as follows. Section~\ref{sec_EKF_EKS} briefly presents the filtering and smoothing DA algorithms used in this work. The main families of methods used in the literature to jointly estimate error covariance matrices $\v{Q}$ and $\v{R}$ are then described. First, moment-based methods are introduced in section~\ref{sec_innovation_statistics}. Then, we describe in section~\ref{sec_likelihood_methods} the likelihood-based methods. We also mention other alternatives in section~\ref{sec_other_methods}, along with methods used in the past but not exactly matching the scope of this review, and diagnostic tools to check the accuracy of $\v{Q}$ and $\v{R}$. Finally, in section~\ref{sec_summary_conclusions_perspectives}, we provide a summary and discussion on what we consider to be the forthcoming challenges in this area.\\

%%%%%%%%%%%%%%%%%%%%%%%%%%%%%%%%%%%%%%%%%%%%%%%%%%%%%%%%%%%%%%%%%%%%%
% FILTERING AND SMOOTHING ALGORITHMS
%%%%%%%%%%%%%%%%%%%%%%%%%%%%%%%%%%%%%%%%%%%%%%%%%%%%%%%%%%%%%%%%%%%%%

\section{Filtering and smoothing algorithms}\label{sec_EKF_EKS}
\black{This review paper focuses on the estimation of $\v{Q}$ and $\v{R}$ in the context of ensemble-based DA methods}. For the overall discussion of the methods and to set the notation, a short description of the ensemble version of the Kalman recursions is presented in this section: the ensemble Kalman filter (EnKF) and ensemble Kalman smoother (EnKS).

The EnKF and EnKS estimate various state vectors $\v{x}^f(k)$, $\v{x}^a(k)$, $\v{x}^s(k)$ and covariance matrices $\v{P}^f(k)$, $\v{P}^a(k)$, $\v{P}^s(k)$, at each time step $1 \le k \le K$, where $K$ represents the total number of assimilation steps. Kalman-based algorithms assume a Gaussian prior distribution $p\left(\v{x}(k)|\v{y}(1:k-1)\right) \sim \mathcal{N}\left(\v{x}^f(k),\v{P}^f(k)\right)$. Then, filtering and smoothing estimates correspond to the Gaussian posterior distributions $p\left(\v{x}(k)|\v{y}(1:k)\right) \sim \mathcal{N}\left(\v{x}^a(k),\v{P}^a(k)\right)$ and $p\left(\v{x}(k)|\v{y}(1:K)\right) \sim \mathcal{N}\left(\v{x}^s(k),\v{P}^s(k)\right)$ of the state conditionally to past/present observations and past/present/future observations respectively.

The basic idea of the EnKF and EnKS is to use an ensemble $\v{x}_1,\dots,\v{x}_{N_e}$ of size $N_e$ to track Gaussian distributions over time with the empirical mean vector $\bar{\v{x}} = \nicefrac{1}{N_e} \sum_{i=1}^{N_e} \v{x}_i$ and the empirical error covariance matrix $\nicefrac{1}{(N_e-1)}\sum_{i=1}^{N_e} \left( \v{x}_i - \bar{\v{x}} \right) \left( \v{x}_i - \bar{\v{x}} \right)^\transp$.
%%% ONLY FOR arXiv VERSION %%%
\newpage
%%% ONLY FOR arXiv VERSION %%%
The EnKF/EnKS equations are divided into three main steps, $\forall i=1,\dots,N_e$ and $\forall k=1,\dots,K$:
\begin{subequations}
\begin{align}
 & \text{\underline{Forecast step (forward in time):}} \nonumber \\
 \v{x}^f_i(k) = & \mathcal{M}_k\left(\v{x}^a_i(k-1)\right) + \eta_i(k) \label{eq_xf} \\
 %\v{P}^f(k) = & \frac{1}{N_e-1} \sum_{i=1}^{N_e} \left( \v{x}^f_{(i)}(k) - \bar{\v{x}}^f(k) \right) \left( \v{x}^f_{(i)}(k) - \bar{\v{x}}^f(k) \right)^\transp \label{eq_Pf} \\
 \nonumber \\
 & \text{\underline{Analysis step (forward in time):}} \nonumber \\
 \v{d}_i(k) = & \v{y}(k) - \mathcal{H}_k\left(\v{x}_i^f(k)\right) + \epsilon_i(k) \label{eq_innov} \\
 \v{K}^f(k) = & \v{P}^f(k) \mathcal{H}_k^\transp \left(\mathcal{H}_k\v{P}^f(k)\mathcal{H}_k^\transp + \v{R}(k)\right)^{-1} \label{eq_kalman_filter_gain} \\
 \v{x}^a_i(k) = & \v{x}^f_i(k) + \v{K}^f(k) \v{d}_i(k) \label{eq_xa} \\
 %\v{P}^a(k) = & \left(\v{I} - \v{K}^f(k) \v{H}_k \right) \v{P}^f(k) \label{eq_Pa} \\
 \nonumber \\
 & \text{\underline{Reanalysis step (backward in time):}} \nonumber \\
 \v{K}^s(k) = & \v{P}^{a}(k) \mathcal{M}_k^\transp \left(\v{P}^f(k+1)\right)^{-1} \label{eq_kalman_smoother_gain} \\
 \v{x}^s_i(k) = & \v{x}^a_i(k) + \v{K}^s(k) \left(\v{x}^s_i(k+1)-\v{x}^f_i(k+1)\right) \label{eq_xs}
 %\v{P}^s(k) = & \v{P}^a(k) \nonumber \\
 %& - \v{K}^s(k) \left( \v{P}^f(k+1) - \v{P}^s(k+1) \right) \v{K}^s(k)^\transp \\
 %\v{P}^{a,f}(k|k+1) = & \v{P}^{a}(k+1) \v{K}^f(k)^\transp \\
 %& + \left( \v{P}^{a,f}(k-1|k) - \v{M} \v{P}^a(k+1)\right) \v{K}^f(k)^\transp
 %\v{P}^s(k,k+1) = & \v{P}^{s}(k+1) \v{K}^s(k)^\transp 
\end{align}
\end{subequations}
with $\v{K}^f(k)$ and $\v{K}^s(k)$ the filter and smoother Kalman gains, respectively. Here, $\v{P}^f(k)$ and $\mathcal{H}_k \v{P}^f(k) \mathcal{H}_k^\transp$ denote the empirical covariance matrices of $\v{x}^f_i(k)$ and $\mathcal{H}_k (\v{x}^f_i(k))$, respectively. Then, $\v{P}^f(k) \mathcal{H}_k^\transp$ and $\v{P}^a(k) \mathcal{M}_k^\transp$ denote the empirical cross-covariance matrices between $\v{x}^f_i(k)$ and $\mathcal{H}_k (\v{x}^f_i(k))$ and between $\v{x}^a_i(k)$ and $\mathcal{M}_k(\v{x}^a_i(k))$, respectively. These quantities are estimated using $N_e$ ensemble members. 

In some of the methods presented in this review, the ensembles are also used to approximate $\mathcal{M}_k$ and $\mathcal{H}_k$ by linear operators $\v{M}_k$ and $\v{H}_k$ such as
\begin{subequations}
\begin{align}
\v{M}_k = & \v{E}^{\mathcal{M}(a)}_k (\v{E}^a_{k-1})^\dagger \label{eq_Mk_linear}\\ \v{H}_k = & \v{E}^{\mathcal{H}(f)}_k (\v{E}^f_k)^\dagger \label{eq_Hk_linear}
\end{align}
\end{subequations}
with $\dagger$ the pseudo-inverse, $\v{E}^{\mathcal{M}(a)}_k$, $\v{E}^a_{k-1}$, $\v{E}^{\mathcal{H}(f)}_k$ and $\v{E}^f_k$ the matrices containing along their columns the ensemble perturbation vectors (the centered ensemble vectors) of $\mathcal{M}_{k}(\v{x}^a_i(k-1))$, $\v{x}^a_i(k-1)$, $\mathcal{H}_k(\v{x}^f_i(k))$ and $\v{x}^f_i(k)$, respectively.

In Eq.~(\ref{eq_innov}), the innovation is denoted as $\v{d}$ and tracked by $\v{d}_1(k), \dots,\v{d}_{N_e}(k)$. The innovation is the key ingredient of the methods presented in sections~\ref{sec_innovation_statistics} and \ref{sec_likelihood_methods}.

%%%%%%%%%%%%%%%%%%%%%%%%%%%%%%%%%%%%%%%%%%%%%%%%%%%%%%%%%%%%%%%%%%%%%
% MOMENT-BASED METHODS
%%%%%%%%%%%%%%%%%%%%%%%%%%%%%%%%%%%%%%%%%%%%%%%%%%%%%%%%%%%%%%%%%%%%%

\section{Moment-based methods\label{sec_innovation_statistics}}

In order to constrain the model and observational errors in DA systems, initial efforts were focused on the statistics of relevant variables which could contain information on covariances. \black{The innovation, given in Eq.~(\ref{eq_innov}), corresponds to the difference between the observations and the forecast in the observation space. This variable implicitly takes into account the $\v{Q}$ and $\v{R}$ covariances.} Unfortunately, as explained in \cite{Blanchet1997AModel}, by using only current observations, their individual contributions cannot be easily disentangled. Thus, the techniques with only the classic innovation \black{$\v{y}(k)-\mathcal{H}_k(\v{x}^f(k))$} are not discussed further in this review.

Two main approaches have been proposed in the literature to address this issue. They are based on the idea of producing multiple equations involving $\v{Q}$ and $\v{R}$. \black{The first one uses different type of innovation statistics (i.e., not only the classic one)}. The second one is based on lag innovations, or differences between consecutive innovations. From a statistical point of view, they refer to the ``methods of moments'', where we construct a system of equations \black{that} links various moments of the innovations with the parameters and then replace the theoretical moments by the empirical ones in these equations.

\subsection{Innovation statistics in the observation space}\label{subsec_desroziers_innovations}

\black{This first approach, based on the Desroziers diagnostic \citep{Desroziers2005DiagnosisSpace}, is historical and now popular in the DA community. It does not exactly fit the topic of this review paper (i.e., estimating the model error $\v{Q}$), since it is based on the inflation of the background covariance matrix $\v{P}^f$. However, this forecast error covariance is defined by $\v{P}^f(k)=\v{M}_k \v{P}^a(k-1) \v{M}_k^\transp + \v{Q}$ in the Kalman filter, considering a linear model operator $\v{M}_k$. Thus, even if DA systems do not use an explicit model error perturbation controlled by $\v{Q}$, the inflation of the background covariance matrix $\v{P}^f$ has similar effects, compensating for the lack of an explicit model uncertainty.}

\cite{Desroziers2005DiagnosisSpace} proposed examining various innovation statistics in the observation space. It is based on different type of innovation statistics between observations, forecasts and analysis, with all of them defined in the observation space: namely, $\v{d}^{o-f}(k) = \v{y}(k) - \mathcal{H}_k\left(\v{x}^f(k)\right)$ as in Eq.~(\ref{eq_innov}) and $\v{d}^{o-a}(k) = \v{y}(k) - \mathcal{H}_k \left(\v{x}^a(k)\right)$. In theory, in the linear and Gaussian case, for unbiased forecast and observation, and when $\v{P}^f(k)$ and $\v{R}(k)$ are correctly specified, the Desroziers innovation statistics should verify the equalities:
\begin{subequations}
\begin{numcases}{}
\mathrm{E}\left[ \v{d}^{o-f}(k) \v{d}^{o-f}(k)^\transp \right] = \v{H}_k \v{P}^f(k) \v{H}_k^\transp + \v{R}(k) \label{equality_innov_1} \\
\mathrm{E}\left[ \v{d}^{o-a}(k) \v{d}^{o-f}(k)^\transp \right] = \v{R}(k) \label{equality_innov_2}
%E \left[ \v{d}^{a-f}(k) \v{d}^{o-f}(k)^\transp \right] = \v{H} \v{P}^f(k) \v{H}^\transp \label{equality_innov_3}
\end{numcases}
\end{subequations}
with $\mathrm{E}$ the expectation operator. Equation~(\ref{equality_innov_1}) is given by using Eq.~(\ref{eq_innov}):
\begin{align}
   \v{d}^{o-f}(k) \v{d}^{o-f}(k)^\transp = & - \v{y}(k) \v{x}^f(k)^\transp \v{H}_k^\transp \nonumber \\
   & - \v{H}_k \v{x}^f(k) \v{y}(k)^\transp \nonumber \\
   & + \v{H}_k \v{x}^f(k) \v{x}^f(k)^\transp \v{H}_k^\transp \nonumber \\ 
   & + \v{y}(k) \v{y}(k)^\transp,
\end{align}
then applying the expectation operator and using the definition of $\v{P}^f$ and $\v{R}$. The observation-minus-forecast innovation statistics in Eq.~(\ref{equality_innov_1}) is not useful to constrain model error $\v{Q}$. Indeed, $\v{d}^{o-f}$ does not depend explicitly on $\v{Q}$, but rather on the forecast error covariance matrix $\v{P}^f$. Thus, the combination of Eq.~(\ref{equality_innov_1}) and Eq.~(\ref{equality_innov_2}) can be used as a diagnosis of the forecast and observational error covariances in the system. A mismatch between the Desroziers statistics and the actual covariances, namely the left- and right-hand side terms in Eq.~(\ref{equality_innov_1}) and Eq.~(\ref{equality_innov_2}), indicates inappropriate estimated covariances $\v{P}^f(k)$ and $\v{R}(k)$.

The forecast covariance $\v{P}^f$ is sometimes badly estimated in ensemble-based assimilation systems. The limitations may be attributed to a number of causes. The limited number of ensemble members produces an over- or, most of the time, underestimation of the forecast variance. \black{Another limitation is the inaccuracies in methods used to sample initial condition or model error.} The underestimation of the forecast covariance produces negative feedback, and the estimated analysis covariance $\v{P}^a$ is thus underestimated, which in turn produces a further underestimation of the forecast covariance in the next cycle. This feedback process leads to filter divergence, as was pointed out by \cite{pham1998singular}, \cite{Anderson1999AForecasts} or \cite{anderson2007adaptive}. To avoid this filter divergence, inflating the forecast covariance $\v{P}^f$ has been proposed. This covariance inflation accounts for both sampling errors and the lack of representation of model errors, like a too small amplitude for $\v{Q}$ or the fact that a bias is omitted in $\gv{\eta}$ and $\gv{\epsilon}$, \black{Eqs.~(\ref{eq_state}) and (\ref{eq_obs})}. In this context, the diagnostics given by the Desroziers innovation statistics have been proposed as a tool to constrain the required covariance inflation in the system.

We distinguish three inflation methods: multiplicative, additive and relaxation-to-prior. In the multiplicative case, the forecast error covariance matrix $\v{P}^f$ is usually multiplied by a scalar coefficient greater than 1 \citep{Anderson1999AForecasts}. Using innovation statistics in the observation space, adaptive procedures to estimate this coefficient have been proposed by \cite{Wang2003ASchemes}, \citet{anderson2007adaptive}, \citet{Anderson2009SpatiallyFilters} conditionally to the spatial location, \cite{Li2009SimultaneousFilter},
\cite{Miyoshi2011TheFilter}, \cite{Bocquet2011EnsembleInflation},
\cite{Bocquet2012CombiningSystems}, \cite{Miyoshi2013EstimatingAssimilation},
\citet{bocquet2015expanding}, \citet{elgharamti2018enhanced}
and \citet{raanes2019apdative}. \black{In order to prevent excessive inflation or deflation, some authors have proposed assuming a priori distribution for the multiplicative inflation factor. The most usual a priori distributions used by the authors are Gaussian in \citet{Anderson2009SpatiallyFilters}, inverse-gamma in \cite{elgharamti2018enhanced} or inverse chi-square in \cite{raanes2019apdative}.}

In practice, multiplicative inflation tends to excessively inflate in the \black{data-sparse} regions and inflate too little in the densely observed regions. As a result, the spread looks like exaggeration of data density (i.e., too much spread in sparsely observed regions, and vice versa). Additive inflation solves this problem, but requires a lot of samples for additive noise; these drawbacks and benefits are discussed in \cite{Miyoshi2010EnsembleSystem}. In the additive inflation case, the diagonal terms of the forecast and analysis empirical covariance matrices is increased \citep{Mitchell2000AnFilter, Corazza2003UseDay,Whitaker2008EnsembleSystem,Houtekamer2009ModelFilter}. This regularization also avoids the problems corresponding to the inversion of the covariance matrices.

The last alternative is the relaxation-to-prior method. In practice, this technique is more efficient than both additive and multiplicative inflations because it maintains a reasonable spread structure. The idea is to relax the reduction of the spread at analysis. We distinguish the method proposed in \cite{Zhang2004ImpactsFilter}, where the forecast and analysis ensemble perturbations are blended, from the one given in \cite{Whitaker2012EvaluatingAssimilation}, which multiplies the analysis ensemble without blending perturbations. This last method is thus a multiplicative inflation, but applied after the analysis, not the forecast. Finally, \cite{Ying2015AnAssimilation} and \cite{Kotsuki2017AdaptiveAtmosphere} proposed methods to adaptively estimate the relaxation parameters using innovation statistics. Their conclusions are that adaptive procedures for relaxation-to-prior methods are robust to sudden changes in the observing networks and observation error settings.

Closely connected to multiplicative inflation estimation is statistical modeling of the error variance terms proposed by \citet{bishop2013a} and \citet{bishop2013b}. From numerical evidence based on the 10-dimensional Lorenz-96 model, the authors assume an inverse-gamma prior distribution for these variances. This distribution allows for an analytic Bayesian update of the variances using the innovations. Building on \cite{Bocquet2011EnsembleInflation,bocquet2015expanding,menetrier2015}, this technique was extended in \cite{satterfield2018} to adaptively tune a mixing ratio between the true and sample variances.

Adaptive covariance inflations are estimation methods directly attached to a traditional filtering method (such as the EnKF used here), with almost negligible overhead computational cost. In practice, the use of this technique does not necessarily imply an additive error term $\gv{\eta}$ in Eq.~(\ref{eq_state}). Thus, it is not a direct estimation of $\v{Q}$ but rather an inflation applied to $\v{P}^f$ in order to compensate for model uncertainties and sampling errors in the EnKFs, as explained in \black{Raanes et al. (2019, their section~4 and appendix C)}. Several DA systems work with an inflation method and use it for its simplicity, low cost, and efficiency. As an example of inflation techniques, the most straightforward inflation estimation is a multiplicative factor $\lambda$ of the incorrectly scaled $\tilde{\v{P}}^f(k)$, so that the corrected forecast covariance is given by $\v{P}^f(k) = \lambda(k) \tilde{\v{P}}^f(k)$. The estimate of the inflation factor is given by taking the trace of Eq.~(\ref{equality_innov_1}):
\begin{equation}\label{lambda_CI}
  \tilde{\lambda}(k) = \frac{\v{d}^{o-f}(k)^\transp \v{d}^{o-f}(k) - \text{Tr} \left( \v{R}(k) \right)}{\text{Tr} \left( \v{H}_k\tilde{\v{P}}^f(k)\v{H}_k^\transp \right)}.
\end{equation}
In practice, the estimated inflation parameter $\tilde{\lambda}$ computed at each time $k$ can be noisy. The use of temporal \black{smoothing of the form} $\lambda(k+1)=\rho\tilde{\lambda}(k)+(1-\rho)\lambda(k)$ is crucial in operational procedures. Alternatively, \cite{Miyoshi2011TheFilter} proposed calculating the estimated variance of $\lambda(k)$, denoted as $\sigma^2_{\lambda(k)}$, using the central limit theorem. Then, $\lambda(k+1)$ is updated using the previous estimate $\lambda(k)$ and the Gaussian distribution with mean $\tilde{\lambda}(k)$ and variance $\sigma^2_{\lambda(k)}$. From the Desroziers diagnostics, at each time step $k$ and when sufficient observations are available, an estimate of \v{R}(k) is possible using Eq.~(\ref{equality_innov_2}). For instance, \cite{Li2009SimultaneousFilter} proposed estimating each component of a diagonal and averaged $\v{R}$ matrix. However, the diagonal terms \black{cannot} take into account spatial correlated error terms, and constant values for observation errors are not realistic in practice. Then, \cite{Miyoshi2013EstimatingAssimilation} proposed additionally estimating the off-diagonal components of the time-dependent matrix $\v{R}(k)$. The \cite{Miyoshi2013EstimatingAssimilation} implementation is summarized in the appendix, Algorithm~\ref{ap_A}.

%Alternatively, \cite{Miyoshi2011TheFilter} proposed augmenting the state vector with the inflation factor whose evolution is governed by a random walk equation such as 
%\begin{equation}
%\label{RW}
%\lambda(k+1)=\tilde{\lambda}(k) + \eta_{\lambda}(k).
%\end{equation}
%In this case, we need to specify an additional parameter for the variance term of this random walk, denoted by $\sigma^2_{\lambda}=Var(\eta_{\lambda} (k))$. This parameter has to be carefully tuned to avoid filter divergence.

The Desroziers diagnostic method has been applied widely to estimate the real observation error covariance matrix $\v{R}$ in Numerical Weather Prediction (NWP). \black{The observations are coming from different sources.} In the case of satellite radiances, \cite{Bormann2010EstimatesData} applied three methods, including the Desroziers diagnostic and the method detailed in \cite{hollingsworth1986statistical} to estimate a constant diagonal term of $\v{R}$ using the innovation $\v{d}^{o-f}$ and its correlations in space, assuming that horizontal correlations in $\v{d}^{o-f}$ samples are purely due to $\v{P}^f$. \cite{weston2014accounting} and \cite{campbell2017accounting} then included the inter-channel observation error correlations of satellite radiances in DA and obtained improved results compared with the case using a diagonal $\v{R}$. For spatial error correlations in $\v{R}$, \cite{kotsuki2017assimilating} estimated the horizontal observation error correlations of satellite-derived precipitation data. Including horizontal observation error correlations in DA for densely-observed data from satellites and radars is more challenging than including inter-channel error correlations in DA. Indeed, the number of horizontally error-correlated observations is much larger, and some recent studies have been tackling this issue (e.g., \cite{guillet2019modelling}).

To conclude, the Desroziers diagnostic is a consistency check and makes it possible to detect if the error covariances $\v{P}^f$ and $\v{R}$ are incorrect. When and how this method can result in accurate or inaccurate estimates, and convergence properties, have been studied in depth by \cite{Waller2016TheoreticalStatistics} and \cite{Menard2016ErrorNetworks}. The Desroziers diagnostic is also useful to estimate off-diagonal terms of $\v{R}$, for instance taking into account the spatial error correlations. \black{However, covariance localization used in the ensemble Kalman filter might induce erroneous estimates of spatial correlations \citep{waller2017diagnosing}.}

\subsection{Lag innovation between consecutive times}\label{subsec_lag_innovations}

Another way to estimate error covariances is to use multiple equations involving $\v{Q}$ and $\v{R}$, exploiting cross-correlations between lag innovations. More precisely, it involves the current innovation $\v{d}(k) = \v{d}^{o-f}(k)$ defined in Eq.~(\ref{eq_innov}) and past innovations $\v{d}(k-1)$, $\dots$, $\v{d}(k-l)$. Lag innovations were introduced by \cite{Mehra1970OnFiltering} to recover $\v{Q}$ and $\v{R}$ simultaneously for Gaussian, linear and stationary dynamic systems. In such a case, $\{\v{d}(k)\}_{k\geq 1}$ is completely characterized by the lagged covariance matrix $\v{C}_{l} = \text{Cov}(\v{d}(k),\v{d}(k-l))$, which is independent of $k$. In other words, the information encoded in $\{\v{d}(k)\}_{k\geq 1}$ is completely equivalent to the information provided by $\{\v{C}_{l}\}_{l\geq 0}$. \black{Moreover, for linear systems in a steady state, analytic relations exist between $\v{Q}$, $\v{R}$ and $\mathrm{E} \left[ \v{d}(k)\v{d}(k-l)^{\transp} \right]$.} However, these linear relations can be dependent and redundant for different lags $l$. Therefore, as stated in \cite{Mehra1970OnFiltering}, only a limited number of $\v{Q}$ components can be recovered.

\cite{Belanger1974EstimationProcess} extended these results to the case of time-varying linear stochastic processes, taking $\v{d}(k)\v{d}(k-l)^{\transp}$ as ``observations'' of $\v{Q}$ and $\v{R}$ and using a secondary Kalman filter to update them iteratively. On the one hand, considering the time-varying case may increase the number of components in $\v{Q}$ that can be estimated. On the other hand, as pointed out in \cite{Belanger1974EstimationProcess}, this method would no longer be analytically exact if $\v{Q}$ and $\v{R}$ were updated adaptively at each time step. One numerical difficulty of B\'elanger's method is that it needs to invert a matrix of size $m^2\times m^2$, where $m$ refers to the dimension of the observation vector. However, this difficulty has been largely overcome by \cite{Dee1985AnSystems} in which the matrix inversion is reduced to $\mathcal{O}(m^3)$, by taking the advantage of the fact that the big matrix comes from some tensor product.

More recent work have focused on high-dimensional and nonlinear systems using the extended or ensemble Kalman filters. \cite{Berry2013AdaptiveSystems} proposed a fast and adaptive algorithm inspired by the use of lag innovations proposed by Mehra. \cite{Harlim2014AnModels} applied the original B\'elanger algorithm empirically to a nonlinear system with sparse observations. \cite{Zhen2015AdaptiveFilters} proposed a modified version of B\'elanger's method, by removing the secondary filter and alternatively solving \v{Q} and \v{R} in a least-squares sense based on the averaged linear relation over a long term.

Here, we briefly describe the algorithm of \cite{Berry2013AdaptiveSystems}, considering the lag-zero and lag-one innovations. The following equations are satisfied in the linear and Gaussian case, for unbiased forecast and observation when $\v{P}^f(k)$ and $\v{R}(k)$ are correctly specified:
\begin{subequations}
\begin{numcases}{}
\mathrm{E}\left[ \v{d}(k) \v{d}(k)^\transp \right] = \v{H}_k \v{P}^f(k) \v{H}_k^\transp + \v{R}(k) \label{lag_innov_0} = \gv{\Sigma}(k) \label{lag_innov_1_Sigma} \\
\mathrm{E}\left[ \v{d}(k) \v{d}(k-1)^\transp \right] = \v{H}_k \v{M}_k \v{P}^f(k-1) \v{H}_{k-1}^\transp \nonumber \\
- \v{H}_k \v{M}_k \v{K}^f(k-1) \gv{\Sigma}(k-1). \label{lag_innov_1}
\end{numcases}
\end{subequations}

Equation~(\ref{lag_innov_1_Sigma}) is equivalent to Eq.~(\ref{equality_innov_1}). Moreover, Eq.~(\ref{lag_innov_1}) results from the fact that developing the expression of $\v{d}(k)$ using consecutively Eqs.~(\ref{eq_obs}), (\ref{eq_state}), (\ref{eq_xf}), and (\ref{eq_xa}), the innovation can be written as
\begin{align}
   \v{d}(k) = & \v{y}(k) - \v{H}_k \v{x}^f(k) \nonumber \\
   = & \v{H}_k \left( \v{x}(k)-\v{x}^f(k) \right) + \gv{\epsilon}(k) \nonumber \\
   = & \v{H}_k \left( \v{M}_k \v{x}(k-1) - \v{x}^f(k) + \gv{\eta}(k) \right) + \gv{\epsilon}(k) \nonumber \\
   = & \v{H}_k \left( \v{M}_k \left(\v{x}(k-1) - \v{x}^a(k-1)\right) + \gv{\eta}(k) \right) + \gv{\epsilon}(k) \nonumber \\
   %= & \v{H}_k \left( \mathcal{M}\left(\v{x}(k-1) - \left( \v{x}^f(k-1) + \v{K}^f(k-1)\v{d}(k-1) \right)\right) + \gv{\eta}(k) \right) + \gv{\epsilon}(k) \nonumber \\
   = & \v{H}_k \v{M}_k \left( \v{x}(k-1) - \v{x}^f(k-1)  - \v{K}^f(k-1)\v{d}(k-1) \right) \nonumber \\
   & + \v{H}_k \gv{\eta}(k) + \gv{\epsilon}(k).
\end{align}
Hence, the innovation product $\v{d}(k) \v{d}(k-1)^\transp$ between two consecutive times is given by
\begin{align}
  & \v{H}_k \v{M}_k \left( \v{x}(k-1) - \v{x}^f(k-1) \right) \v{d}(k-1)^\transp \nonumber \\
  & - \v{H}_k \v{M}_k \left( \v{K}^f(k-1)\v{d}(k-1) \right) \v{d}(k-1)^\transp \nonumber \\
  & + \v{H}_k \gv{\eta}(k) \v{d}(k-1)^\transp + \gv{\epsilon}(k) \v{d}(k-1)^\transp,
\end{align}
and assuming that the model $\gv{\eta}$ and observation $\gv{\epsilon}$ error noises are white and mutually uncorrelated, then $\mathrm{E}\left[\gv{\eta}(k)\v{d}(k-1)^\transp\right]=0$ and $\mathrm{E}\left[\gv{\epsilon}(k)\v{d}(k-1)^\transp\right]=0$. Finally, developing $\mathrm{E}\left[ \v{d}(k) \v{d}(k-1)^\transp \right]$, Eq.~(\ref{lag_innov_1}) is satisfied.

%\begin{equation}
%\v{d}(k) \v{d}(k-1)^\transp = \left[ \v{H}_k \left( \v{M}_k \left(\v{x}(k-1) - \v{x}^f(k-1)  - \v{K}^f(k-1)\v{d}(k-1) \right) + \gv{\eta}(k) \right) + \gv{\epsilon}(k) \right] \v{d}(k-1)^\transp
%\end{equation}

The algorithm in \cite{Berry2013AdaptiveSystems} is summarized in the appendix, Algorithm~\ref{ap_B}. It is based on an adaptive estimation of $\v{Q}(k)$ and $\v{R}(k)$, which satisfies the following relations in the linear and Gaussian case:
\begin{subequations}
\begin{align}
 \tilde{\v{P}}(k) = & \left( \v{H}_k \v{M}_k \right)^{-1} \v{d}(k) \v{d}(k-1)^\transp \v{H}_{k-1}^{-\transp}, \nonumber \\
         & + \v{K}^f(k-1) \v{d}(k-1) \v{d}(k-1)^\transp \v{H}_{k-1}^{-\transp} \label{eq_P_berry} \\
 \tilde{\v{Q}}(k) = & \tilde{\v{P}}(k) - \v{M}_{k-1} \v{P}^a(k-2) \v{M}_{k-1}^{\transp}, \label{eq_Q_berry} \\
 \tilde{\v{R}}(k) = & \v{d}(k) \v{d}(k)^\transp - \v{H}_k \v{P}^f(k) \v{H}_k^{\transp}. \label{eq_R_berry}
\end{align}
\end{subequations}

In operational applications, when the number of observations is not equal to the number of components in state $\v{x}$, $\v{H}$ is not a square matrix and Eq.~(\ref{eq_P_berry}) is ill-defined. To avoid the inversion of $\v{H}$, \cite{Berry2013AdaptiveSystems} proposed considering parametric models for $\v{Q}$ and then solving a linear system associated with \black{Eqs.~(\ref{eq_P_berry}) and (\ref{eq_Q_berry})}. It is written as a least-squares problem such that
\begin{align}
 \tilde{\v{Q}}(k) = & \argmin_\v{Q} || \v{d}(k) \v{d}(k-1)^\transp \nonumber \\
 & + \v{H}_k \v{M}_k \v{K}^f(k-1) \v{d}(k-1) \v{d}(k-1)^\transp \nonumber \\
 & - \v{H}_k \v{M}_k \v{M}_{k-1} \v{P}^a(k-2) \v{M}_{k-1}^\transp  \v{H}_{k-1}^\transp \nonumber \\
 & - \v{H}_k \v{M}_k \v{Q} \v{H}_{k-1}^\transp ||. \label{eq_ls_Q_berry}
\end{align}

In this adaptive procedure, joint estimations of $\tilde{\v{Q}}(k)$ and $\tilde{\v{R}}(k)$ can abruptly vary over time. Thus, the temporal smoothing of the covariances being estimated becomes crucial. As suggested by \cite{Berry2013AdaptiveSystems}, such temporal smoothing between current and past estimates is a reasonable choice:
\begin{subequations}
\begin{align}
 \v{Q}(k+1) & = \rho \tilde{\v{Q}}(k) + (1-\rho) \v{Q}(k), \label{eq_LI_tau_Q}\\
 \v{R}(k+1) & = \rho \tilde{\v{R}}(k) + (1-\rho) \v{R}(k) \label{eq_LI_tau_R}
\end{align}
\end{subequations}
with $\v{Q}(1)$ and $\v{R}(1)$ the initial conditions and $\rho$ the smoothing parameter. When $\rho$ is large (close to 1), weight is given to the current estimates $\tilde{\v{Q}}$ and $\tilde{\v{R}}$, and when $\rho$ is small (close to 0) it gives smoother $\v{Q}$ and $\v{R}$ sequences. The value of $\rho$ \black{is arbitrary and} may depend on the system and how it is observed. For instance, in the case where the number of observations equals the size of the system, \cite{Berry2013AdaptiveSystems} uses \black{$\rho=5 \times 10^{-5}$} in order to estimate the full matrix \v{Q} for the Lorenz-96 model.

The algorithm in \cite{Berry2013AdaptiveSystems} only considers lag-zero and lag-one innovations. By incorporating more lags, \cite{Zhen2015AdaptiveFilters} and \cite{Harlim2018EnsembleFilters} showed that it makes it possible to deal with the case in which some components of Q are not identifiable from the method in \cite{Berry2013AdaptiveSystems}. For instance, let us consider the \black{two-dimensional} system with any stationary operator $\v{M}$ and $\v{H}=[ 1 , 0 ]$, meaning that only the first component of the system is observed. This is a linear, Gaussian, stationary system, and Mehra's theory implies that \black{two} parameters of $\v{Q}$ are identifiable. \black{However, using only lag-one innovations as in \cite{Berry2013AdaptiveSystems}, Eq.~(\ref{eq_ls_Q_berry}) becomes a scalar equation and only \black{one} parameter of $\v{Q}$ can be determined.} The idea of considering more lag innovations to estimate more components of $\v{Q}$ was tested in \cite{Zhen2015AdaptiveFilters}. Numerical results show that considering more than one lag can improve the estimates of \black{$\v{Q}$ and $\v{R}$}. \black{For instance, \cite{Zhen2015AdaptiveFilters} focused on the Lorenz-96 model. Results show that when $\v{Q}$ is stationary, the trace of $\v{Q}$ and $\v{R}$ are equal, and when observations are taken at twenty fixed equally spaced grid points for every five integration time steps, the optimal RMSE of the estimates of $\v{Q}$ and $\v{R}$ is achieved when four time lags are considered. But with more lags, the performance is degraded.}

To summarize, methods based on lag innovation between consecutive times have been studied for a long time in the signal processing community. The original methods \citep{Mehra1970OnFiltering,Belanger1974EstimationProcess} were analytically established for linear systems with Gaussian noises. Inspired by these foundational ideas, empirical methods have been established for nonlinear systems in DA \citep{Berry2013AdaptiveSystems, Harlim2014AnModels, Zhen2015AdaptiveFilters}. Although these methods have not been tested in any operational experiment, the idea of using lagged innovations seems to have significant potential.

%%%%%%%%%%%%%%%%%%%%%%%%%%%%%%%%%%%%%%%%%%%%%%%%%%%%%%%%%%%%%%%%%%%%%
% LIKELIHOOD-BASED METHODS
%%%%%%%%%%%%%%%%%%%%%%%%%%%%%%%%%%%%%%%%%%%%%%%%%%%%%%%%%%%%%%%%%%%%%

\section{Likelihood-based methods}\label{sec_likelihood_methods}

This section focuses on methods based on the likelihood of the observations, given a set of statistical parameters. The conceptual idea behind what we refer to as likelihood-based methods is to determine the optimal statistical parameters (i.e., $\v{Q}$ and $\v{R}$) that maximize the likelihood function for a given set of observations which may be distributed over time. In this way, the aim is to derive estimation methods that use the observations to find the most suitable, or most likely parameters.

Early studies in \black{\cite{Dee1995On-lineAssimilation}, \cite{Blanchet1997AModel}, \cite{Mitchell2000AnFilter} and \cite{Liang2012MaximumAssimilation}} proposed finding the optimal $\v{Q}$ and $\v{R}$ that maximize the current innovation likelihood at time $k$. Unfortunately, if only the current observations are used, the joint estimation of $\v{Q}$ and $\v{R}$ is not well constrained \citep{Todling2015AMethod}. To tackle this issue, several solutions have been recently proposed where the likelihood function considers observations distributed in time over several assimilation cycles.

The likelihood-based methods are broadly divided into two categories. One approach uses a Bayesian framework. It assumes a priori knowledge about the parameters and estimate jointly the posterior distribution of $\v{Q}$ and $\v{R}$ together with the state of the system, or alternatively to estimate them in a two-stage process\footnote{Some of the methods presented in section~\ref{sec_innovation_statistics} also use the Bayesian philosophy \black{;} for instance they assume a priori distribution for the multiplicative inflation parameter $\lambda$ \citep{Anderson2009SpatiallyFilters,elgharamti2018enhanced}.}. The second one is based on the frequentist viewpoint and attempts a point estimate of the parameters by maximizing a total likelihood function.

\subsection{Bayesian inference}

In the Bayesian framework, the elements of the covariance matrices $\v{Q}$ and $\v{R}$ are assumed to have a priori distributions which are controlled by hyperparameters. In practice, it is difficult to have prior distributions for each element of $\v{Q}$ and $\v{R}$, especially for large DA systems. Instead, parametric forms are used for the matrices, typically describing the shape and level noise. We denote the corresponding parameters as $\gv{\theta}$. 

The inference in the Bayesian framework aims to determine the posterior density $p \left( \gv{\theta}|\v{y}(1:k) \right)$. Two techniques have appeared, the first based on a state augmentation and the second based on a rigorous Bayesian update of the posterior distribution.

\subsubsection{State augmentation}
In the Bayesian framework, $\gv{\theta}$ is a random variable such that the state is augmented with these parameters by defining $\v{z}(k)= \left( \v{x}(k),\gv{\theta} \right)$. To define an augmented state-space model, one has to define an evolution equation for the parameters. This leads to a new state-space model of the form of \black{Eqs.~(\ref{eq_state}) and (\ref{eq_obs})} with $\v{x}$ replaced by $\v{z}$. Therefore, the state and the parameters are estimated jointly using the DA algorithms.

State augmentation was first proposed in \cite{schmidt1966applications} and is known as the Schmidt--Kalman filter. This technique was mainly used to estimate both the state of the system and additional parameters, including bias, forcing terms and physical parameters. These kinds of parameters are strongly related to the state of the system \citep{Ruiz2013EstimatingReview}. Therefore, they are identifiable and suitable for an augmented state approach. However, \cite{Stroud2007SequentialFilter} and later \cite{Delsole2010StateModels} formally demonstrated that augmentation methods fail for variance parameters like $\v{Q}$ and $\v{R}$. \black{The explanation is that in the EnKF, the empirical forecast covariance $\v{P}^f$ is computed using all the ensemble members, each one with a different realization of the random variable $\gv{\theta}$. Thus, $\v{P}^f$ and consequently the Kalman gain $\v{K}^f$, are mixing the effects of $\v{Q}$ and $\v{R}$ parameters contained in $\gv{\theta}$. Therefore, after applying Eq.~(\ref{eq_xa}), the update of $\v{z}$ corresponding to the $\gv{\theta}$ parameters is the same for all the parameters.} To capture the impact of a single variance parameter on the prediction covariance and circumvent the limitation of the state augmentation, \black{\cite{scheffler2018inference} proposed} to use an ensemble of states integrated with the same variance parameter. The choice of an ensemble of states for each variance parameter leads to two nested ensemble Kalman filters. The technique performs successfully under different model error covariance structures but has an important computational cost.

Another critical aspect of state augmentation is that one needs to define an evolution model for the augmented state $\v{z}(k)=\left( \v{x}(k),\gv{\theta}(k) \right)$. If persistence is assumed in the parameters such that they are constant in time, this leads to filter degeneracy, since the estimated variance of the error in $\theta$ is bound to decrease in time. To prevent or at least mitigate this issue, it was suggested to use an independent inflation factor on the parameters \citep{Ruiz2013EstimatingTreatment} or to impose artificial stochastic dynamics for $\gv{\theta}$, typically a random walk or AR(1) model, as introduced in Eq.~(\ref{eq_AR1}) and proposed in \cite{liu2001combined}. The tuning of the parameters introduced in these artificial dynamics may be difficult in practice, and this introduces bias into the procedure, which is hard to quantify.

\subsubsection{Bayesian update of the posterior distribution}

Instead of the inference of the joint posterior density using a state augmentation strategy, the state $\v{x}(k)$ and parameters $\gv{\theta}$ can be divided into a two-step inference procedure using the following formula:
\begin{align}
& p\left(\v{x}(k),\gv{\theta}|\v{y}(1:k)\right) = & \nonumber \\ 
& p\left(\v{x}(k)|\v{y}(1:k),\gv{\theta}\right) p\left(\gv{\theta}|\v{y}(1:k)\right), & \label{eq_bayesian}
\end{align}
which is a direct consequence of the conditional density definition. In Eq.~(\ref{eq_bayesian}), $p\left(\v{x}(k)|\v{y}(1:k),\gv{\theta}\right)$ represents the posterior distribution of the state, given the observations and the parameter $\gv{\theta}$. It can be computed using a filtering DA algorithm. The second term on the right-hand side of Eq.~(\ref{eq_bayesian}) corresponds to the posterior distribution of the parameters, given the observations up to time $k$. The latter can be updated sequentially using the following Bayesian hierarchy:
\begin{align}
& p\left(\gv{\theta}|\v{y}(1:k)\right) \propto & \nonumber \\ 
& p\left(\v{y}(k)|\v{y}(1:k-1),\gv{\theta}\right) p\left(\gv{\theta}|\v{y}(1:k-1)\right),  & \label{eq_bayesian2}
\end{align}
where $p\left(\v{y}(k)|\v{y}(1:k-1),\gv{\theta}\right)$ is the likelihood of the innovations.

Different approximations have been used for $p\left(\gv{\theta}|\v{y}(1:k)\right)$ in Eq.~(\ref{eq_bayesian2}) \black{;} these include parametric models based on Gaussian \citep{Stroud2018AEstimation}, inverse-gamma \citep{Stroud2007SequentialFilter} or Wishart distributions \citep{Ueno2016BayesianFilters}, particle-based approximations \citep{Frei2012SequentialFilters, Stroud2018AEstimation} and grid-based approximation \citep{Stroud2018AEstimation}.

The methods proposed in the literature also differ by the approximation used for the likelihood of the innovations. We \black{emphasize} that $p\left(\v{y}(k)|\v{y}(1:k-1),\gv{\theta}\right)$ needs to be evaluated for different values of $\gv{\theta}$ at each time step, and that this requires applying the filter from the initial time with a single value of $\gv{\theta}$, which is computationally impossible for applications in high dimensions. To reduce computational time, it is generally assumed that $\v{x}^f$ and $\v{P}^f$ are independent of $\gv{\theta}$, and only observations $\v{y}(k-l:k-1)$ in a small time window from the current observation are used when computing the likelihood of the innovations (see \cite{Ueno2016BayesianFilters,Stroud2018AEstimation} for a more detailed discussion). A summary of the Bayesian method from \cite{Stroud2018AEstimation} is given in the appendix, Algorithm~\ref{ap_C}. It was implemented within the EnKF framework and is one of the most recent studies based on the Bayesian approach.

\black{Applications of the Bayesian methodology in the DA context are now discussed}. It has mainly been used to estimate shape and noise parameters of $\v{Q}$ and $\v{R}$ error covariance matrices. For instance, \cite{Purser2003AAssimilation} and \cite{Solonen2014EstimatingFiltering} estimated statistical parameters controlling the magnitude of the variance and the spatial dependencies in the model error $\v{Q}$, assuming that $\v{R}$ is known. There are also applications aimed at estimating parameters governing the shape of the observation error covariance matrix $\v{R}$ only: \cite{Frei2012SequentialFilters} and \cite{Stroud2018AEstimation} in the Lorenz-96 system, \cite{Winiarek2012EstimationPlant,Winiarek2014EstimationObservations} for the inversion of the source term of airborne radionuclides using a regional atmospheric model, and \cite{Ueno2016BayesianFilters} using a shallow-water model to assimilate satellite altimetry.

As pointed out in \cite{Stroud2007SequentialFilter}, Bayesian update algorithms work best when the number of unknown parameters in $\gv{\theta}$ is small. This \black{limitation} may explain why the joint estimation of parameters controlling both model and observation error covariances is not systematically addressed. For instance, \cite{Stroud2007SequentialFilter} used the EnKF with the Lorenz-96 model for the estimation of a common multiplicative scalar parameter for predefined matrices $\v{Q}$ and $\v{R}$. Alternatively, \cite{Stroud2018AEstimation} tested the Bayesian method on different spatio-temporal systems to estimate the signal-to-noise ratio between $\v{Q}$ and $\v{R}$. Nevertheless, based on the experiments about the importance of the signal-to-noise ratio $\norm{\v{P}^f}/\norm{\v{R}}$ presented in Fig.~\ref{fig_1}, we know that this \black{estimation of the ratio} is not optimal.

Widely used in the statistical community, the Bayesian framework is useful incorporating physical knowledge about error covariance matrices and constraining their estimation process. In the DA literature, authors have used a priori distributions for the shape and noise parameters of $\v{Q}$ or $\v{R}$, but rarely both. In practice, only a limited number of parameters can be estimated. To address this issue, \cite{Stroud2007SequentialFilter} suggested combining Bayesian algorithms with other techniques.

\subsection{Maximization of the total likelihood.}\label{subsec_maxtotlik}

The innovation likelihood at time $k$, $p\left(\v{y}(k)|\v{y}(1:k-1),\gv{\theta}\right)$ in Eq.~(\ref{eq_bayesian2}), can be maximized to find the optimal $\gv{\theta}$ \black{(i.e., $\v{Q}$ and $\v{R}$ matrices or parameterizations of them)}. But in practice, when this maximization is done at each time step, two issues arise. Firstly, the innovation covariance matrix $\gv{\Sigma}(k) = \v{H}_k \v{P}^f(k) \v{H}_k^\transp + \v{R}(k)$ combines the information about $\v{R}$ and $\v{Q}$, the latter being contained in $\v{P}^f$. When using only time $k$, it is difficult to disentangle the model and observation error covariances; in practice, the aforementioned studies only estimated one of them. Secondly, the number of observations at each time step is in general limited and, as pointed out by \cite{Dee1995On-lineAssimilation}, available observations should exceed ``the number of tunable parameters by two or three orders of magnitude''. To overcome these limitations, a reasonable alternative is to use a batch of observations within a time window and to assume $\gv{\theta}$ to be constant in time. The resulting total likelihood expressed sequentially through conditioning is given by
\begin{equation}
p\left(\v{y}(1:K)|\gv{\theta}\right) = \prod_{k=1}^K p\left(\v{y}(k)|\v{y}(1:k-1),\gv{\theta}\right). \label{eq_tot_lik_incomplete}
\end{equation}
\black{Because} it is an integration of innovation likelihoods over a long period of time from $k=1$ to $k=K$, Eq.~(\ref{eq_tot_lik_incomplete}) provides more observational information to estimate $\v{Q}$ and $\v{R}$. The maximization of this total likelihood has been applied for the estimation of deterministic and stochastic parameters (related to $\v{Q}$) using a direct sequential optimization procedure \citep{Delsole2010StateModels}. \cite{Ueno2010MaximumModel} used a grid-based procedure to estimate noise levels and spatial correlation lengths of $\v{Q}$ and a noise level for $\v{R}$. This grid-based method uses predefined sets of covariance parameters and evaluates the different combinations to find the one that maximizes the likelihood criterion. \cite{Brankart2010EfficientSignals} also proposed a method using the same criterion but adding (at the initial time) information on scale and correlation length parameters of $\v{Q}$ and $\v{R}$. This information is only given the first time, and is progressively forgotten over time, using a decreasing exponential factor. The marginalization of the hidden state in Eq.~(\ref{eq_tot_lik_incomplete}) considers all the previous observations, and in practice it requires the use of a filter. The maximization of the total likelihood $p\left(\v{y}(1:K)|\gv{\theta}\right)$ to estimate model error covariance $\v{Q}$ was conducted in \cite{Pulido2018StochasticMethods}, where they used a gradient-based optimization technique and the EnKF.

The likelihood function given in Eq.~(\ref{eq_tot_lik_incomplete}) only depends on the observations $\v{y}$. This likelihood can be written in a different way, taking into account both the observations and the hidden state $\v{x}$. Indeed, the marginalization of the hidden state to obtain the total likelihood can be produced using the whole trajectory of the state from $k=0$ to the last time step $K$ all at once. It is given by
\begin{equation}
p(\v{y}(1:K)|\gv{\theta}) = \int p(\v{x}(0:K),\v{y}(1:K)|\gv{\theta}) \mathrm{d} \v{x}(0:K). \label{comp-lik}
\end{equation}

In practice, the maximization of the total likelihood as a function of statistical parameters $\gv{\theta}$ is not possible, since the total likelihood cannot be evaluated directly, nor its gradient with regard to the parameters \citep{Pulido2018StochasticMethods}. \cite{Shumway1982AnAlgorithm} proposed using an iterative procedure based on the \black{expectation--maximization} algorithm (hereinafter denoted as EM). They applied it to estimate the parameters of a linear state-space model, with linear dynamics\black{, and} a linear observational operator and Gaussian errors. The EM algorithm was introduced by \cite{Dempster1977MaximumAlgorithm}.

Each iteration of the EM algorithm consists of two steps. In the expectation step (E-step), the posterior density $p(\v{x}(0:K)|\v{y}(1:K),\gv{\theta}_{(n)})$ is determined conditioned on the batch of observations $\v{y}(1:K)$ and given the parameters $\gv{\theta}_{(n)}=\left(\v{Q}_{(n)}, \v{R}_{(n)}\right)$ from the previous iteration or initial guess. In practice, this is obtained through the application of a smoother like the EnKS. Then, the M-step relies on the maximization of an intermediate function, depending on the posterior density obtained in the E-step. The intermediate function is defined by the conditional expectation
\begin{equation}
%\mathcal{Q}\left(\theta|\theta_{n-1}\right) = 
\mathrm{E}\left[\log \left( p(\v{x}(0:K),\v{y}(1:K)|\gv\theta) \right) | \v{y}(1:K), \theta_{(n)}\right].
\label{interFn}
\end{equation}

If as in \black{Eqs.~(\ref{eq_state}) and (\ref{eq_obs})} the observational and model errors are assumed to be additive, unbiased and Gaussian, the expression for the logarithm of the joint density in Eq.~(\ref{interFn}) is given by
\begin{align}
  - & \frac{1}{2} \{ \sum_{k=1}^K \lVert \v{x}(k) - \mathcal{M}(\v{x}(k-1)) \rVert_{\v Q}^2 +  \log |\v{Q}| \nonumber \\
+ & \lVert \v{y}(k) - \mathcal{H}(\v{x}(k)) \rVert_{\v R}^2 + \log |\v{R}| \}+ c \label{logJoint}
\end{align}
where $\lVert\v v\rVert_{\v A}^2$ is defined to be equal to $\v v^\transp \v A^{-1} \v v$ and $c$ is a constant independent of $\v{Q}$ and $\v{R}$. In this case, an analytic expression for the optimal error covariances at each iteration of the EM algorithm can be obtained. The estimators of the parameters that maximize Eq.~(\ref{interFn}) using Eq.~(\ref{logJoint}) are
\begin{subequations}
\begin{align}
    \v{Q}_{(n+1)} = & \frac{1}{K} \sum_{k=1}^K \mathrm{E} [ (\v{x}(k) - \mathcal{M}(\v{x}(k-1))) \nonumber \\ 
    & (\v{x}(k) - \mathcal{M}(\v{x}(k-1)))^\transp  | \v{y}(1:K), \theta_{(n)} ] \label{Qest}
\end{align}
and
\begin{align}
    \v{R}_{(n+1)} = & \frac{1}{K} \sum_{k=1}^K  \mathrm{E} [ (\v{y}(k) - \mathcal{H}(\v{x}(k))) \nonumber \\
    & (\v{y}(k) - \mathcal{H}(\v{x}(k)))^\transp | \v{y}(1:K), \theta_{(n)} ]. \label{Rest}
\end{align}
\end{subequations}

The application of the EM algorithm for the estimation of \v{Q} and \v{R} is rather straightforward. Starting from $\v{Q}_{(1)}$ and $\v{R}_{(1)}$, an ensemble Kalman smoother is applied with this first guess and the batch of observations $\v{y}(1:K)$ to obtain the posterior density $p(\v{x}(0:K)|\v{y}(1:K),\gv{\theta}_{(1)})$. Then \black{Eqs.~(\ref{Qest}) and (\ref{Rest})} are used to update and obtain $\v{Q}_{(2)}$ and $\v{R}_{(2)}$. Next, a new application of the smoother is conducted using the parameters $\v{Q}_{(2)}$ and $\v{R}_{(2)}$ and the observations $\v{y}(1:K)$, the new resulting states are used in \black{Eqs.~(\ref{Qest}) and (\ref{Rest})} to estimate $\v{Q}_{(3)}$ and $\v{R}_{(3)}$, and so on. As a diagnostic of convergence or as a stop criterion, the product of innovation likelihood functions given in Eq.~(\ref{eq_tot_lik_incomplete}) is evaluated using a filter. The EM algorithm guarantees that the total likelihood increases in each iteration and that the sequence $\gv{\theta}_{(n)}$ converges to a local maximum \citep{Wu1983OnAlgorithm}. A summary of the EM method (using EnKF and EnKS) from \cite{Dreano2017EstimatingAlgorithm} is given in the appendix, Algorithm~\ref{ap_D}.

EM is a well-known algorithm used in the statistical community. This procedure is parameter-free and robust, due to the large number of observations used to approximate the likelihood when using a long batch period \citep{Shumway1982AnAlgorithm}. Although the use of the EM algorithm is still limited in DA, \black{it is becoming more and more popular}. Some studies have implemented the EM algorithm for estimating only the observation error matrix $\v{R}$. For instance, \cite{Ueno2014IterativeFilters} used \black{the model proposed in \cite{zebiak1987model}} and satellite altimetry observations, whereas \cite{Liu2017UncertaintyRadionuclides} used an air quality model for accidental pollutant source retrieval. But the estimation of only the observation error covariance is limited, and other studies have tried to jointly estimate model error $\v{Q}$ and $\v{R}$ matrices, for instance as in \cite{tandeo2015offline} for an orographic subgrid-scale nonlinear observation operator. Then, \cite{Dreano2017EstimatingAlgorithm} and \cite{Pulido2018StochasticMethods} used the EM procedure to produce joint estimation of $\v{Q}$ and $\v{R}$ matrices in the Lorenz-63 and stochastic parameters of the Lorenz-96 systems, respectively. \black{Recently, \cite{yang2019estimation} extended the EM procedure for the estimation of physical parameters in a one-dimensional shallow water model, more specifically for the identification of stochastic subgrid terms.} \black{Lastly, an online adaptation of the EM algorithm for the estimation of $\v{Q}$ and $\v{R}$ at each time step, after the filtering procedure, has been proposed in \cite{cocucci2020model}. In this adaptive case, the likelihood is averaged locally over time, see \cite{Cappe2011OnlineExpectation-Maximisation} for more details.}

To our knowledge, EM has not been tested yet on operational systems with large observation- and state-space. In that case, the use of parametric forms for the matrices $\v{Q}$ and $\v{R}$ is essential to reduce the number of statistical parameters $\gv{\theta}$ to estimate. For instance, \cite{Dreano2017EstimatingAlgorithm} and \cite{Liu2017UncertaintyRadionuclides} showed that in the particular cases where covariances are diagonal or of the form $\alpha \v{A}$ with $\v{A}$ a positive definite matrix, expressions in Eq.~(\ref{Qest}) and Eq.~(\ref{Rest}) are simplified, and a suboptimal $\gv{\theta}$ in the space of the parametric covariance form \black{can be} obtained.

%%%%%%%%%%%%%%%%%%%%%%%%%%%%%%%%%%%%%%%%%%%%%%%%%%%%%%%%%%%%%%%%%%%%%
% OTHER METHODS
%%%%%%%%%%%%%%%%%%%%%%%%%%%%%%%%%%%%%%%%%%%%%%%%%%%%%%%%%%%%%%%%%%%%%

\section{Other methods}\label{sec_other_methods}
In this section, we describe other methods that have been used to estimate $\v{Q}$ and $\v{R}$, and that cannot be included in the categories presented in the previous sections. In particular, we report here about methods that are applied either a posteriori, after DA cycles, or without applying any DA algorithms.

\subsection{Analysis (or reanalysis) increment approach}\label{subsec_analysis_increment}

\black{This first method is based on previous DA outputs.} The key idea here is to use the analysis (or reanalysis) increments to provide a realistic sample of model errors from which statistical moments, such as the covariance matrix $\v{Q}$, can be empirically estimated. This assumes that the sequence of reanalysis $\v{x}^s$ (or analysis $\v{x}^a$) is the best available representation of the true process $\v{x}$. In that case, the following approximation in Eq.~(\ref{eq_state}) is made:
\begin{align}
 \gv{\eta}(k) = & \mathcal{M}\left(\v{x}(k-1)\right)-\v{x}(k) \nonumber \\
 \approx & \mathcal{M}\left(\v{x}^s(k-1)\right)-\v{x}^s(k).\label{eq_state_approx}
\end{align}
In this approximation, it is implicitly assumed that the estimated state is the truth, so that the initial condition at time $k-1$ is neglected. A similar approximation of the true process by $\v{x}^a$ or $\v{x}^s$ in Eq.~(\ref{eq_obs}) can be used to estimate the observation error covariance matrix $\v{R}$.

In practice, the analysis (or reanalysis) increment method is applied after a DA filter (or smoother) to estimate the $\v{Q}$ matrix. This method was originally introduced by \cite{Leith1978ObjectivePrediction}, and later used to account for model error in the context of ensemble Kalman filters, using analysis and reanalysis increments by \cite{Mitchell2015AccountingFiltering}. Along this line, \cite{rodwell2007using} also proposed evaluating the average of instantaneous analysis increments to represent the systematic forecast tendencies of a model.

\subsection{Covariance matching}\label{subsec_covariance matching}
The covariance matching method was introduced by \cite{Fu1993FittingModel}. It involves matching sample covariance matrices to their theoretical expectations. Thus, it is a method of moments, similar to the work in \cite{Desroziers2005DiagnosisSpace}, except that covariance matching is performed on a set of historical observations and numerical simulations (noted $\v{x}^{sim}$), without applying any DA algorithms. It has been extended by \cite{Menemenlis2000ErrorData} to time-lagged innovations, as first considered in \cite{Belanger1974EstimationProcess}.

In the case of a constant and linear observation operator $\v{H}$, the basic idea in \cite{Fu1993FittingModel} is to assume the following system
\begin{subequations}
\begin{numcases}{}
\v{x}^{sim}(k) = \v{x}(k) + \gv{\eta}^{sim}(k), \label{eq_state_bis}\\
\gv{\eta}^{sim}(k) = \v{A} \gv{\eta}^{sim}(k-1) + \gv{\eta}(k), \label{eq_eta_bis}\\
\v{H}\v{x}^{sim}(k) - \v{y}(k) = \v{H} \gv{\eta}^{sim}(k) + \gv{\epsilon}(k), \label{eq_obs_bis}
\end{numcases}
\end{subequations}
with $\v{A}$ a transition matrix close to the identity matrix, assuming slow variations of the numerical simulation errors (noted $\gv{\eta}^{sim}$). In Eq.~(\ref{eq_eta_bis}) and Eq.~(\ref{eq_obs_bis}), the definitions of $\gv{\eta}$ and $\gv{\epsilon}$ errors remain similar, as in the general \black{Eqs.~(\ref{eq_state}) and (\ref{eq_obs})}.

Assuming that $\v{Q}$ and $\v{R}$ are constant over time, $\gv{\epsilon}$ is uncorrelated from $\v{x}$ and from $\gv{\eta}^{sim}$, then Eq.~(\ref{eq_obs_bis}) and Eq.~(\ref{eq_state_bis}) yield to the following estimates of $\v{R}$ and $\v{P}^{sim}$ (the latter represents the error covariance of the numerical simulations):
\begin{subequations}
\begin{align}
\widehat{\v{R}} = & \frac{1}{2} \{ \mathrm{E} [ (\v{y} - \v{H} \v{x}^{sim}) (\v{y} - \v{H} \v{x}^{sim})^\transp ] \nonumber \\
& - \mathrm{E} [ (\v{H} \v{x}^{sim}) (\v{H} \v{x}^{sim})^\transp] + \mathrm{E} [ \v{y} \v{y}^\transp ]\},\\
\v{H}\widehat{\v{P}}^{sim}\v{H}^\transp = & \frac{1}{2} \{ \mathrm{E} [ (\v{y} - \v{H} \v{x}^{sim}) (\v{y} - \v{H} \v{x}^{sim})^\transp ] \nonumber \\
& + \mathrm{E} [ (\v{H} \v{x}^{sim}) (\v{H} \v{x}^{sim})^\transp ] - \mathrm{E} [ \v{y} \v{y}^\transp ]\}. \label{eq_HPsimHt}
\end{align}
\end{subequations}
where $\mathrm{E}$ is the expectation operator over time. Then, an estimate of $\v{Q}$ is obtained using Eq.~(\ref{eq_eta_bis}), Eq.~(\ref{eq_HPsimHt}) and assuming that $\v{P}^{sim}$ has a unique time-invariant limit.

\subsection{Forecast sensitivity}
In operational meteorology, it is critical to learn the sensitivity of the forecast accuracy to various parameters of a DA system, in particular the error statistics of both the model and the observations. This is why a significant portion of literature considers the tuning problem of $\v{R}$ and $\v{Q}$ through the lens of the sensitivity of the forecast to these parameters. The computation of those sensitivities can be seen as a first-order correction or diagnostic for such an estimation. The forecast sensitivities are computed either using the adjoint model \citep{daescu2010,daescu2013} in the context of variational methods, or a forecast ensemble \citep{hotta2017} in the context of the EnKF.

The basic idea is to compute at each assimilation cycle an innovation between forecast and analysis, noted $\v{d}^{f-a}(k) = \v{x}^f(k) - \v{x}^a(k)$. Then, the forecast sensitivity is given by $\v{d}^{f-a}(k)^\transp \v{S} \v{d}^{f-a}(k)$ with $\v{S}$ a diagonal scaling matrix, to normalize the components of $\v{d}^{f-a}$. $\v{Q}$ and $\v{R}$ estimates are the matrices that minimize $\v{d}^{f-a}(k)$. The adjoint or the ensemble are thus used to compute the partial derivatives of this forecast sensitivity. w.r.t. $\v{Q}$ and $\v{R}$.

\section{Conclusions and perspectives}\label{sec_summary_conclusions_perspectives}

As often considered in data assimilation, this review paper also deals with model and observation errors that are assumed additive and Gaussian with covariance matrices $\v{Q}$ and $\v{R}$. The model error corresponds to the dynamic model deficiencies
to represent the underlying physics, whereas the observation error corresponds to the instrumental noise and the representativity error. Model and observation errors are assumed to be uncorrelated and white in time. The model and observations are also assumed unbiased, a strong assumption for real data assimilation applications.

The discussion starts with the aid of an illustration of the individual and joint impacts of improperly calibrated covariances using a linear toy model. The experiments clearly showed that to achieve reasonable filter accuracy (i.e., in terms of root mean squared error), it is crucial to carefully define both $\v{Q}$ and $\v{R}$. The effect on the coverage probability of a \black{mis-specification} of $\v{Q}$ or $\v{R}$ is also highlighted. This coverage probability is related to the estimated covariance of the reconstructed state, and thus to the uncertainty quantification in data assimilation. After the \black{one-dimensional} illustration, the core of the paper gives an overview of various methods to jointly estimate the $\v{Q}$ and $\v{R}$ error covariance matrices: they are summarized and compared below.

\subsection{Comparison of existing methods for estimating $\v{Q}$ and $\v{R}$}

We mainly focused in this review on four methodologies for the joint estimation of the error covariances $\v{Q}$ and $\v{R}$. The methods are summarized in Table~\ref{tab_summary}. They correspond to classic estimation methods, based on statistical moments or likelihoods. The main difference between the four methods comes from the innovations taken into account: the total innovation, as in the EM algorithm proposed by \cite{Shumway1982AnAlgorithm}; lag innovations, following the idea given in \cite{Mehra1970OnFiltering}; or different type of innovations in the observation space, as in \cite{Desroziers2005DiagnosisSpace}. Additionally, to constrain the estimation, hierarchical Bayesian approaches use prior distributions for the shape parameters of $\v{Q}$ and $\v{R}$.

\begin{table*}[!ht]
\begin{small}
\caption{Comparison of several methods to estimate error covariance matrices $\v{Q}$ and $\v{R}$ in data assimilation.}\label{tab_summary}
\begin{center}
\begin{tabular}{p{2.7cm} p{3.1cm} p{2cm} p{2.6cm} p{3.3cm}}
\topline
\centering{Estimation method} & \centering{Criteria} & \centering{Estimation of covariance $\v{Q}$} & \centering{\black{Suitable for non-Gaussian errors}} & \centering{Application to the highest complexity model} \tabularnewline
\midline
Method of moments & Innovation statistics in the observation space & No (inflation \black{of $\v{P}^f$} instead) & \black{No} & NWP\\
%\hline
Method of moments & Lag innovation between consecutive times & Yes & \black{No} & Lorenz-96\\
%\hline
Likelihood methods & \black{Bayesian update of the posterior distribution} & No (or joint parameter with $\v{R}$) & \black{Yes (using particle filters, not EnKF)} & Shallow water\\
%\hline
Likelihood methods & Maximization of the total likelihood & Yes & \black{Yes (using particle filters, not EnKF)} & Two-scale Lorenz-96\\
\botline
\end{tabular}
\end{center}
\end{small}
\end{table*}

\black{Most of the methods estimate the model error $\v{Q}$. The exception is the one using the Desroziers diagnostic, dealing with different type of innovations in the observation space, which instead estimates an inflation factor for $\v{P}^f$.} Moreover, the methods are mainly defined online, meaning that they aim to estimate $\v{Q}$ and $\v{R}$ adaptively, together with the current state of the system. Consequently, these methods require additional tunable parameters to smooth the estimated covariances over time. \black{However, most of the methods presented in this review also have an offline variant. In that case, a batch of observations is used to estimate $\v{Q}$ and $\v{R}$. In some methods, such as the EM algorithm, the parameters are determined iteratively. These offline approaches avoid the use of additional smoothing parameters.}

\black{Throughout this review paper, as usually stated in DA, it is assumed that model error $\gv{\eta}$ and observation error $\gv{\epsilon}$, defined in Eqs.~(\ref{eq_state}) and (\ref{eq_obs}), are Gaussian. Consequently, the distribution of the innovations are also Gaussian. The four presented methods use this property to build estimates of $\v{Q}$ and $\v{R}$ adequately. But, if $\gv{\eta}$ and $\gv{\epsilon}$ are non-Gaussian, Desroziers diagnostic and lag-innovation methods are not suitable anymore. However, the EM procedures and Bayesian methods are still relevant, although they must be used with an appropriate filter (e.g., particle filters), not Kalman-based algorithms (i.e., assuming a Gaussian distribution of the state). Recently, the treatment of non-Gaussian error distributions in DA has been explored in \cite{katzfuss2019ensemble}, using hierarchical state-space models. This Bayesian framework allows to handle unknown variables that cannot be easily included in the state vector (e.g., parameters of $\v{Q}$ and $\v{R}$) and to model non-Gaussian observations.}

The four methods have been applied at different levels of complexity. \black{For instance, Bayesian inference methods (due to their algorithm complexity) and the EM algorithm (due to its computational cost) have so far only been applied to small dynamic models. However, the online version of the EM algorithm is less consuming and opens new perspectives of applications on larger models.} On the other hand, methods using innovation statistics in the observation space have already been applied to NWP models.

The four methods summarized in Table~\ref{tab_summary} show differences in maturity in terms of applications and methodological aspects. This review also shows that there are still remaining challenges and possible improvements for the four methods.

\subsection{Remaining challenges for each method}

The first challenge concerns the improvements of adaptive techniques regarding additional parameters that control the variations of $\v{Q}$ and $\v{R}$ estimates over time. Instead of using fixed values for these parameters, for instance fixed $\rho$ in the lag innovations or $\sigma^2_\lambda$ in the inflation methods, we suggest using time-dependent adaptations. This \black{adaptive solution could} avoid the problems of instabilities close to the solution. Another option could be to adapt these procedures, working with stable parameter values (small $\rho$, low $\sigma^2_\lambda$) and iterating the procedures on a batch of observations, as in the EM algorithm. This \black{offline variant} was suggested and tested in \cite{Desroziers2005DiagnosisSpace} with encouraging results. To the best of our knowledge, it has not yet been tested with lag-innovation methods.

\black{The second challenge concerns considering time-varying error covariance matrices. The adaptive procedures, based on online estimations with temporal smoothing of $\v{Q}$ and $\v{R}$, are supposed to capture slowly evolving covariances. On the contrary, offline methods like the EM algorithm are working on a batch of observations, assuming that $\v{Q}$ and $\v{R}$ are constant over the batch period. Online solutions for the EM algorithm, with the likelihood averaged locally over time \citep{cocucci2020model}, could also capture slow evolution of the covariances. Another simple solution could be to work on small sets of observations, named as mini-batches, and to apply the EM algorithm in each set using the previous estimates as an initial guess. These intermediate schemes are of common use in machine learning.}

\black{A third challenge has to do with the assumption, used by all of the methods described herein, that observation and model errors are mutually independent. Nevertheless, as pointed out in \cite{Berry2018CorrelationAssimilation}, observation and model error are often correlated in real data assimilation problems (e.g., for satellite retrieval of Earth observations that uses model outputs in the inversion process). Methods based on Bayesian inference can, in principle, exploit existing model-to-observation correlations by using a prior joint distribution (i.e., not two individual ones). The explicit taking into account of this correlation can then constrain the optimization procedure. This is not possible in the other approaches described in this review, at least not in their standard known formulations, and the presence of model-observation correlation can deteriorate their accuracy.}

\black{A fourth challenge is common to all the methods presented in this review. Iterative versions of the presented algorithms need initial values or distributions for $\v{R}$ and $\v{Q}$ (or $\v{B}=\v{P}^f$ in the case of Desroziers). But, as mentioned in \cite{Waller2016TheoreticalStatistics} for the Desrorziers diagnostics, there is no guarantee that the algorithms will converge to the optimal solution. Indeed, in such an optimization problem, there are possibly several local and non-optimal solutions. Bad specifications of $\v{R}$, $\v{Q}$, or $\v{B}$ in the initial DA cycle will affect the final estimation results. There are several solutions to avoid this convergence problem: initialize the covariance matrices using physical expertise, execute the iterative algorithms several times with different initial covariance matrices, or use stochastic perturbations in the optimization algorithms to avoid to be trapped in local solutions. These aspects of convergence and sensitivity to initial conditions have so far been poorly addressed. It is therefore necessary to check which method is robust in practice.}

The last remaining challenge concerns the estimation of other statistical parameters of the state-space model given in \black{Eqs.~(\ref{eq_state}) and (\ref{eq_obs})} and associated filters. Indeed, the initial conditions $\v{x}(0)$ and $\v{P}(0)$ are crucial for certain satellite retrieval problems and have to be estimated. This is the case, for instance, when the time sequence of observations is short (i.e., shorter than the spinup time of the filter with an uninformative prior) or when filtering and smoothing are repeated on various iterations, as in the EM algorithm. Estimation methods should also consider the estimation of systematic or time-varying biases, the deterministic part of $\gv{\eta}$ and $\gv{\epsilon}$. This was initially proposed by \cite{Dee1999Maximum-LikelihoodMethodology} and tested in \cite{Dee1999Maximum-LikelihoodApplications} in the case of maximizing the innovation likelihood, in \cite{Dee2005BiasAssimilation} in a state augmentation formulation, and was adapted to a Bayesian update formulation in \cite{Liu2017UncertaintyRadionuclides} and in \cite{Berry2017CorrectingAssimilation}. \black{Recently, the joint estimation of bias and covariance error terms, for the treatment of brightness temperatures from the European geostationary satellite, has been successfully applied in \cite{Merchant2020BiasReferences}.}

\subsection{Perspectives for geophysical DA}

Beyond the aforementioned potential improvements in the existing techniques, specific research directions need to be taken by the data assimilation community. The main one concerns the realization of a comprehensive numerical evaluation of the different methods for the estimation of $\v{Q}$ and $\v{R}$, built on an agreed experimental framework and a consensus model. Such an effort would help to evaluate (i) the pros and cons of the different methods (including their capability to deal with high dimensionality, localization in ensemble methods\black{, and} their practical feasibility), (ii) their effects on different error statistics (RMSE, coverage probabilities\black{, and} other diagnostics), (iii) the potential combination of the various methods (especially those considering constant or adaptive covariances), and (iv) the capability to take into account other sources of error (due for instance to improper parameterizations, multiplicative errors\black{, or} forcing terms).

\black{The use of a realistic DA problem, with a high-dimensional state-space and a limited and heterogeneous observational coverage should be addressed in the future. In that realistic case, the observational information per degree of freedom will be significantly lower, and the estimates of $\v{Q}$ and $\v{R}$ will deteriorate. Parametric versions of these error covariance matrices will therefore be necessary. Among the parameters, some of them will control the variances, and will be different depending on the variable. Other parameters will control the spatial correlation lengths, that could be isotropic or anisotropic, depending on the region of interest and the considered variable. Cross-correlations between variables will also have to be considered. Consequently, $\v{Q}$ and $\v{R}$ will be block-matrices with as few parameters as possible.}

A further challenge for future work is the evaluation of the feasibility of estimating non-additive, non-Gaussian, and time-correlated noises under the current estimation frameworks. In this way, the need for observational constraints for the stochastic perturbation methods in the NWP community could be considered within the estimation framework discussed in this review.

%%%%%%%%%%%%%%%%%%%%%%%%%%%%%%%%%%%%%%%%%%%%%%%%%%%%%%%%%%%%%%%%%%%%%
% ACKNOWLEDGMENTS
%%%%%%%%%%%%%%%%%%%%%%%%%%%%%%%%%%%%%%%%%%%%%%%%%%%%%%%%%%%%%%%%%%%%%

\acknowledgments{This work has been carried out as part of the Copernicus Marine Environment Monitoring Service (CMEMS) 3DA project. CMEMS is implemented by Mercator Ocean in the framework of a delegation agreement with the European Union. This work was also partially supported by FOCUS Establishing Supercomputing Center of Excellence. CEREA is a member of Institut Pierre Simon Laplace (IPSL). A.~C. has been funded by the project REDDA (\#250711) of the Norwegian Research Council. \black{A.~C. was also supported by the Natural Environment Research Council (Agreement PR140015 between NERC and the National Centre for Earth Observation).} We thank Paul Platzer, a second-year PhD student, who helped to popularize the summary and the introduction, and John C. Wells, Gilles-Olivier Gu\'egan and Aim\'ee Johansen for their English grammar corrections. We also thank the \black{five} anonymous \black{reviewers} for their precious comments and ideas to improve this review paper. Finally, we are immensely grateful to Prof. David M. Schultz, Editor in Chief of the \textit{Monthly Weather Review}, for his detailed advice and careful reading of the paper.

%%%%%%%%%%%%%%%%%%%%%%%%%%%%%%%%%%%%%%%%%%%%%%%%%%%%%%%%%%%%%%%%%%%%%
% APPENDIXES
%%%%%%%%%%%%%%%%%%%%%%%%%%%%%%%%%%%%%%%%%%%%%%%%%%%%%%%%%%%%%%%%%%%%%

\newpage
\appendix[]
\appendixtitle{Four main algorithms to jointly estimate $\v{Q}$ and $\v{R}$ in data assimilation}

\begin{algorithm}[h]%[H]
- initialize inflation factor (for instance $\lambda(1)=1$)\;
 \For{$k$ in 1:$K$}{
  \For{$i$ in 1:$N_e$}{
  - compute forecast $\v{x}^f_i(k)$ using Eq.~(\ref{eq_xf})\;
  - compute innovation $\v{d}_i(k)$ using Eq.~(\ref{eq_innov})\;
  }
  - compute empirical covariance $\tilde{\v{P}}^f(k)$ of the $\v{x}^f_i(k)$\;
  - compute $\v{K}^f(k)$ using Eq.~(\ref{eq_kalman_filter_gain}) where $\tilde{\v{P}}^f(k) \mathcal{H}_k^\transp$ and $\mathcal{H}_k \tilde{\v{P}}^f(k) \mathcal{H}_k^\transp$ are inflated by $\lambda(k)$\;
  \For{$i$ in 1:$N_e$}{
  - compute analysis $\v{x}^a_i(k)$ using Eq.~(\ref{eq_xa})\;
  }
  - compute mean innovations $\v{d}^{o-f}(k)$ and $\v{d}^{o-a}(k)$ with $\v{d}^{o-f}_i(k) = \v{y}(k) - \mathcal{H}_k (\v{x}^f_i(k))$ and $\v{d}^{o-a}_i(k) = \v{y}(k) - \mathcal{H}_k ( \v{x}^a_i(k) )$\;
  - update $\v{R}(k)$ from Eq.~(\ref{equality_innov_2}) using the cross-covariance between $\v{d}^{o-f}_i(k)$ and $\v{d}^{o-a}_i(k)$\;
  - estimate $\tilde{\lambda}(k)$ using Eq.~(\ref{lambda_CI}) where $\mathcal{H}_k \tilde{\v{P}}^f(k) \mathcal{H}_k^\transp$ is inflated by $\lambda(k)$\;
  - update $\lambda(k+1)$ using temporal smoother\;
 }
 \caption{Adaptive algorithm for the EnKF \citep{Miyoshi2013EstimatingAssimilation} } \label{ap_A}
\end{algorithm}
%\nicefrac{1}{N_e}\sum_{i=1}^{N_e}

\newpage
\begin{algorithm}[h]%[H]
- initialize $\v{Q}(1)$ and $\v{R}(1)$\;
 \For{$k$ in 1:$K$}{
  \For{$i$ in 1:$N_e$}{
  - compute forecast $\v{x}^f_i(k)$ using Eq.~(\ref{eq_xf})\;
  - compute innovation $\v{d}_i(k)$ using Eq.~(\ref{eq_innov})\;
  }
  - compute $\v{K}^f(k)$ using Eq.~(\ref{eq_kalman_filter_gain})\;
  \For{$i$ in 1:$N_e$}{
  - compute analysis $\v{x}^a_i(k)$ using Eq.~(\ref{eq_xa})\;
  }
  - apply Eq.~(\ref{eq_P_berry}) to get $\tilde{\v{P}}(k)$ using linearizations of $\v{M}_k$ and $\v{H}_k$ given in \black{Eqs.~(\ref{eq_Mk_linear}) and (\ref{eq_Hk_linear})}\;
  - estimate $\tilde{\v{Q}}(k)$ using Eq.~(\ref{eq_Q_berry})\;
  - estimate $\tilde{\v{R}}(k)$ using Eq.~(\ref{eq_R_berry})\;
  - update $\v{Q}(k+1)$ and $\v{R}(k+1)$ using temporal smoothers\;
 }
 \caption{Adaptive algorithm for the EnKF \citep{Berry2013AdaptiveSystems}} \label{ap_B}
\end{algorithm}

\newpage
\begin{algorithm}[h]%[H]
- define a priori distributions for $\gv{\theta}$ (shape parameters of $\v{Q}$ and $\v{R}$)\;
 \For{$k$ in 1:$K$}{
  \For{$i$ in 1:$N_e$}{
  - draw samples $\gv{\theta}_i(k)$ from $p\left(\gv{\theta}|\v{y}(1:k-1)\right)$\;
  - compute forecast $\v{x}^f_i(k)$ using Eq.~(\ref{eq_xf}) with $\gv{\theta}_i(k)$\;
  - compute innovation $\v{d}_i(k)$ using Eq.~(\ref{eq_innov}) with $\gv{\theta}_i(k)$\;
  }
  - compute $\v{K}^f(k)$ using Eq.~(\ref{eq_kalman_filter_gain})\;
  \For{$i$ in 1:$N_e$}{
  - compute analysis $\v{x}^a_i(k)$ using Eq.~(\ref{eq_xa})\;
  }
- approximate Gaussian likelihood of innovations $p\left(\v{y}(k)|\v{y}(1:k-1),\gv{\theta}(k)\right)$ using empirical mean $\bar{\v{d}}(k) = \frac{1}{N_e}\sum_{i=1}^{N_e} \v{d}_i(k)$ and empirical covariance $\gv{\Sigma}(k) = \frac{1}{N_e-1} \sum_{i=1}^{N_e} \left( \v{d}_i(k) - \bar{\v{d}}(k) \right) \left( \v{d}_i(k) - \bar{\v{d}}(k) \right)^\transp$ with $ \v{d}_i(k) = \v{y}(k) - \mathcal{H}_k (\v{x}^f_i(k))$\;
  %- approximate Gaussian likelihood of innovations $p\left(\v{y}(k)|\v{y}(1:k-1),\gv{\theta}(k)\right)$ with $\gv{\theta}(k)$ the sample mean of $\gv{\theta}_i(k)$\;
  - update $p\left(\gv{\theta}|\v{y}(1:k)\right)$ using Eq.~(\ref{eq_bayesian2})\;
 }
 \caption{Adaptive algorithm for the EnKF \citep{Stroud2018AEstimation}} \label{ap_C}
\end{algorithm}

\newpage
\begin{algorithm}[h]%[H]
- define $\gv{\theta}$ (shape parameters of $\v{Q}$ and $\v{R}$)\;
- set $p\left(\v{y}(1:K)|\gv{\theta}_{(0)}\right)=+\infty$\;
- initialize $n=1$, $\gv{\theta}_{(1)}$ and $\epsilon$ (stop condition)\;
\While{$p\left(\v{y}(1:K)|\gv{\theta}_{(n)}\right) - p\left(\v{y}(1:K)|\gv{\theta}_{(n-1)}\right) > \epsilon$}{
 \For{$k$ in 1:$K$}{
  \For{$i$ in 1:$N_e$}{
  - compute forecast $\v{x}^f_i(k)$ using Eq.~(\ref{eq_xf})\;
  - compute innovation $\v{d}_i(k)$ using Eq.~(\ref{eq_innov})\;
  }
  - compute $\v{K}^f(k)$ using Eq.~(\ref{eq_kalman_filter_gain})\;
  \For{$i$ in 1:$N_e$}{
  - compute analysis $\v{x}^a_i(k)$ using Eq.~(\ref{eq_xa})\;
  }
 }
 \For{$k$ in $K$:1}{
  - compute $\v{K}^s(k)$ using Eq.~(\ref{eq_kalman_smoother_gain})\;
  \For{$i$ in 1:$N_e$}{
  - compute reanalysis $\v{x}^s_i(k)$ using Eq.~(\ref{eq_xs})\;
  }
 }
 - increment $n\leftarrow n+1$\;
 - estimate $\v{Q}_{(n)}$ using Eq.~(\ref{Qest})\;
 - estimate $\v{R}_{(n)}$ using Eq.~(\ref{Rest})\;
}
 \caption{EM algorithm for the EnKF/EnKS \citep{Dreano2017EstimatingAlgorithm}} \label{ap_D}
\end{algorithm}

%%%%%%%%%%%%%%%%%%%%%%%%%%%%%%%%%%%%%%%%%%%%%%%%%%%%%%%%%%%%%%%%%%%%%
% REFERENCES
%%%%%%%%%%%%%%%%%%%%%%%%%%%%%%%%%%%%%%%%%%%%%%%%%%%%%%%%%%%%%%%%%%%%%

\newpage
\bibliographystyle{ametsoc2014}
\bibliography{Mendeley}

\begin{thebibliography}{107}
\providecommand{\natexlab}[1]{#1}
\providecommand{\url}[1]{\texttt{#1}}
\renewcommand{\UrlFont}{\rmfamily}
\providecommand{\urlprefix}{URL }
\expandafter\ifx\csname urlstyle\endcsname\relax
  \providecommand{\doi}[1]{doi:\discretionary{}{}{}#1}\else
  \providecommand{\doi}{doi:\discretionary{}{}{}\begingroup
  \urlstyle{rm}\Url}\fi
\providecommand{\eprint}[2][]{\url{#2}}

\bibitem[{Anderson(2007)}]{anderson2007adaptive}
Anderson, J.~L., 2007: {An adaptive covariance inflation error correction
  algorithm for ensemble filters}. \textit{Tellus A: Dynamic Meteorology and
  Oceanography}, \textbf{59~(2)}, 210--224.

\bibitem[{Anderson(2009)}]{Anderson2009SpatiallyFilters}
Anderson, J.~L., 2009: {Spatially and temporally varying adaptive covariance
  inflation for ensemble filters}. \textit{Tellus, Series A: Dynamic
  Meteorology and Oceanography}, \textbf{61~(1)}, 72--83.

\bibitem[{Anderson and Anderson(1999)Anderson, and
  Anderson}]{Anderson1999AForecasts}
Anderson, J.~L., and S.~L. Anderson, 1999: {A Monte Carlo Implementation of the
  Nonlinear Filtering Problem to Produce Ensemble Assimilations and Forecasts}.
  \textit{Monthly Weather Review}, \textbf{12~(127)}, 2741--2758.

\bibitem[{B{\'{e}}langer(1974)}]{Belanger1974EstimationProcess}
B{\'{e}}langer, P.~R., 1974: {Estimation of noise covariance matrices for a
  linear time-varying stochastic process}. \textit{Automatica},
  \textbf{10~(3)}, 267--275.

\bibitem[{Berry and Harlim(2017)Berry, and
  Harlim}]{Berry2017CorrectingAssimilation}
Berry, T., and J.~Harlim, 2017: {Correcting Biased Observation Model Error in
  Data Assimilation}. \textit{Monthly Weather Review}, \textbf{145~(7)},
  2833--2853.

\bibitem[{Berry and Sauer(2013)Berry, and Sauer}]{Berry2013AdaptiveSystems}
Berry, T., and T.~Sauer, 2013: {Adaptive ensemble Kalman filtering of
  non-linear systems}. \textit{Tellus, Series A: Dynamic Meteorology and
  Oceanography}, \textbf{65~(20331)}, 1--16.

\bibitem[{Berry and Sauer(2018)Berry, and
  Sauer}]{Berry2018CorrelationAssimilation}
Berry, T., and T.~Sauer, 2018: {Correlation between system and observation
  errors in data assimilation}. \textit{Monthly Weather Review},
  \textbf{146~(9)}, 2913--2931.

\bibitem[{Bishop and Satterfield(2013)Bishop, and Satterfield}]{bishop2013a}
Bishop, C.~H., and E.~A. Satterfield, 2013: {Hidden error variance theory. Part
  I: Exposition and analytic model}. \textit{Monthly Weather Review},
  \textbf{141~(5)}, 1454--1468.

\bibitem[{Bishop et~al.(2013)Bishop, Satterfield,, and Shanley}]{bishop2013b}
Bishop, C.~H., E.~A. Satterfield, and K.~T. Shanley, 2013: {Hidden error
  variance theory. Part II: An instrument that reveals hidden error variance
  distributions from ensemble forecasts and observations}. \textit{Monthly
  Weather Review}, \textbf{141~(5)}, 1469--1483.

\bibitem[{Blanchet et~al.(1997)Blanchet, Frankignoul,, and
  Cane}]{Blanchet1997AModel}
Blanchet, I., C.~Frankignoul, and M.~A. Cane, 1997: {A comparison of adaptive
  kalman filters for a tropical pacific ocean model}. \textit{Monthly Weather
  Review}, \textbf{125~(1)}, 40--58.

\bibitem[{Bocquet(2011)}]{Bocquet2011EnsembleInflation}
Bocquet, M., 2011: {Ensemble Kalman filtering without the intrinsic need for
  inflation}. \textit{Nonlinear Processes in Geophysics}, \textbf{18~(5)},
  735--750.

\bibitem[{Bocquet et~al.(2015)Bocquet, Raanes,, and
  Hannart}]{bocquet2015expanding}
Bocquet, M., P.~N. Raanes, and A.~Hannart, 2015: {Expanding the validity of the
  ensemble Kalman filter without the intrinsic need for inflation}.
  \textit{Nonlinear Processes in Geophysics}, \textbf{22}, 645--662.

\bibitem[{Bocquet and Sakov(2012)Bocquet, and
  Sakov}]{Bocquet2012CombiningSystems}
Bocquet, M., and P.~Sakov, 2012: {Combining inflation-free and iterative
  ensemble Kalman filters for strongly nonlinear systems}. \textit{Nonlinear
  Processes in Geophysics}, \textbf{19}, 383--399.

\bibitem[{Bormann et~al.(2010)Bormann, Collard,, and
  Bauer}]{Bormann2010EstimatesData}
Bormann, N., A.~Collard, and P.~Bauer, 2010: {Estimates of spatial and
  interchannel observation-error characteristics for current sounder radiances
  for numerical weather prediction. II: Application to AIRS and IASI data}.
  \textit{Quarterly Journal of the Royal Meteorological Society},
  \textbf{136~(649)}, 1051--1063.

\bibitem[{Brankart et~al.(2010)Brankart, Cosme, Testut, Brasseur,, and
  Verron}]{Brankart2010EfficientSignals}
Brankart, J.-M., E.~Cosme, C.-E. Testut, P.~Brasseur, and J.~Verron, 2010:
  {Efficient adaptive error parameterizations for square root or ensemble
  Kalman filters: Application to the control of ocean mesoscale signals}.
  \textit{Monthly Weather Review}, \textbf{138~(3)}, 932--950.

\bibitem[{Buehner(2010)}]{buehner2010error}
Buehner, M., 2010: {Error statistics in data assimilation: Estimation and
  modelling}. \textit{Data Assimilation}, Springer, 93--112.

\bibitem[{Campbell et~al.(2017)Campbell, Satterfield, Ruston,, and
  Baker}]{campbell2017accounting}
Campbell, W.~F., E.~A. Satterfield, B.~Ruston, and N.~L. Baker, 2017:
  {Accounting for correlated observation error in a dual-formulation 4D
  variational data assimilation system}. \textit{Monthly Weather Review},
  \textbf{145~(3)}, 1019--1032.

\bibitem[{Capp{\'{e}}(2011)}]{Cappe2011OnlineExpectation-Maximisation}
Capp{\'{e}}, O., 2011: {Online Expectation-Maximisation}. \textit{Mixtures:
  Estimation and Applications}, Wiley Series in Probability and Statistics,
  1--53.

\bibitem[{Carrassi et~al.(2018)Carrassi, Bocquet, Bertino,, and
  Evensen}]{Carrassi2018DataPerspectives}
Carrassi, A., M.~Bocquet, L.~Bertino, and G.~Evensen, 2018: {Data Assimilation
  in the Geosciences: An overview on methods, issues and perspectives}.
  \textit{WIREs Clim Change}, \textbf{9~(5)}, e535.

\bibitem[{Chapnik et~al.(2004)Chapnik, Desroziers, Rabier,, and
  Talagrand}]{Chapnik2004PropertiesAssimilation}
Chapnik, B., G.~Desroziers, F.~Rabier, and O.~Talagrand, 2004: {Properties and
  first application of an error-statistics tuning method in variational
  assimilation}. \textit{Quarterly Journal of the Royal Meteorological
  Society}, \textbf{130~(601)}, 2253--2275.

\bibitem[{Cocucci et~al.(2020)Cocucci, Pulido, Lucini,, and
  Tandeo}]{cocucci2020model}
Cocucci, T.~J., M.~Pulido, M.~Lucini, and P.~Tandeo, 2020: {Model error
  covariance estimation in particle and ensemble Kalman filters using an online
  expectation-maximization algorithm}. \textit{arXiv preprint
  arXiv:2003.02109}.

\bibitem[{Corazza et~al.(2003)Corazza, Kalnay, Patil, Morss, Cai, Szunyogh,
  Hunt,, and Yorke}]{Corazza2003UseDay}
Corazza, M., E.~Kalnay, D.~J. Patil, R.~Morss, M.~Cai, I.~Szunyogh, B.~R. Hunt,
  and J.~A. Yorke, 2003: {Use of the breeding technique to estimate the
  structure of the analysis “errors of the day”}. \textit{Nonlinear
  Processes in Geophysics}, \textbf{10~(3)}, 233--243.

\bibitem[{Daescu and Langland(2013)Daescu, and Langland}]{daescu2013}
Daescu, D.~N., and R.~H. Langland, 2013: {Error covariance sensitivity and
  impact estimation with adjoint 4D-Var: theoretical aspects and first
  applications to NAVDAS-AR}. \textit{Quarterly Journal of the Royal
  Meteorological Society}, \textbf{139}, 226--241.

\bibitem[{Daescu and Todling(2010)Daescu, and Todling}]{daescu2010}
Daescu, D.~N., and R.~Todling, 2010: {Adjoint sensitivity of the model forecast
  to data assimilation system error covariance parameters}. \textit{Quarterly
  Journal of the Royal Meteorological Society}, \textbf{136}, 2000--2012.

\bibitem[{Daley(1991)}]{Daley1991AtmosphericAnalysis}
Daley, R., 1991: {Atmospheric data analysis}. Cambridge University Press, 457
  pp.

\bibitem[{Daley(1992)}]{Daley1992EstimatingAssimilation}
Daley, R., 1992: {Estimating Model-Error Covariances for Application to
  Atmospheric Data Assimilation}. \textit{Monthly Weather Review},
  \textbf{120~(8)}, 1735--1746.

\bibitem[{Dee(1995)}]{Dee1995On-lineAssimilation}
Dee, D.~P., 1995: {On-line estimation of error covariance parameters for
  atmospheric data assimilation}. \textit{Monthly Weather Review},
  \textbf{123~(4)}, 1128--1145.

\bibitem[{Dee(2005)}]{Dee2005BiasAssimilation}
Dee, D.~P., 2005: {Bias and data assimilation}. \textit{Quarterly Journal of
  the Royal Meteorological Society}, \textbf{131~(613)}, 3323--3343.

\bibitem[{Dee et~al.(1985)Dee, Cohn, Dalcher,, and Ghil}]{Dee1985AnSystems}
Dee, D.~P., S.~E. Cohn, A.~Dalcher, and M.~Ghil, 1985: {An efficient algorithm
  for estimating noise covariances in distributed systems}. \textit{IEEE
  Transactions on Automatic Control}, \textbf{30~(11)}, 1057--1065.

\bibitem[{Dee et~al.(1999{\natexlab{a}})Dee, Gaspari, Redder, Rukhovets,, and
  da~Silva}]{Dee1999Maximum-LikelihoodMethodology}
Dee, D.~P., G.~Gaspari, C.~Redder, L.~Rukhovets, and A.~M. da~Silva,
  1999{\natexlab{a}}: {Maximum-likelihood estimation of forecast and
  observation error covariance parameters. Part I: Methodology}.
  \textit{Monthly Weather Review}, \textbf{127~(1992)}, 1822--1834.

\bibitem[{Dee et~al.(1999{\natexlab{b}})Dee, Gaspari, Redder, Rukhovets,, and
  da~Silva}]{Dee1999Maximum-LikelihoodApplications}
Dee, D.~P., G.~Gaspari, C.~Redder, L.~Rukhovets, and A.~M. da~Silva,
  1999{\natexlab{b}}: {Maximum-likelihood estimation of forecast and
  observation error covariance parameters. Part II: Applications}.
  \textit{Monthly Weather Review}, \textbf{127~(1992)}, 1835--1849.

\bibitem[{Delsole and Yang(2010)Delsole, and Yang}]{Delsole2010StateModels}
Delsole, T., and X.~Yang, 2010: {State and parameter estimation in stochastic
  dynamical models}. \textit{Physica D: Nonlinear Phenomena},
  \textbf{239~(18)}, 1781--1788.

\bibitem[{Dempster et~al.(1977)Dempster, Laird,, and
  Rubin}]{Dempster1977MaximumAlgorithm}
Dempster, A.~P., N.~M. Laird, and D.~B. Rubin, 1977: {Maximum Likelihood from
  Incomplete Data via the EM Algorithm}. \textit{Journal of the Royal
  Statistical Society. Series B (Methodological)}, \textbf{39~(1)}, 1--38.

\bibitem[{Desroziers et~al.(2005)Desroziers, Berre, Chapnik,, and
  Poli}]{Desroziers2005DiagnosisSpace}
Desroziers, G., L.~Berre, B.~Chapnik, and P.~Poli, 2005: {Diagnosis of
  observation, background and analysis-error statistics in observation space}.
  \textit{Quarterly Journal of the Royal Meteorological Society},
  \textbf{131~(613)}, 3385--3396.

\bibitem[{Desroziers and Ivanov(2001)Desroziers, and
  Ivanov}]{Desroziers2001DiagnosisAssimilation}
Desroziers, G., and S.~Ivanov, 2001: {Diagnosis and adaptive tuning of
  observation-error parameters in a variational assimilation}.
  \textit{Quarterly Journal of the Royal Meteorological Society},
  \textbf{127~(574)}, 1433--1452.

\bibitem[{Dreano et~al.(2017)Dreano, Tandeo, Pulido, Chonavel, AIt-El-Fquih,,
  and Hoteit}]{Dreano2017EstimatingAlgorithm}
Dreano, D., P.~Tandeo, M.~Pulido, T.~Chonavel, B.~AIt-El-Fquih, and I.~Hoteit,
  2017: {Estimating model error covariances in nonlinear state-space models
  using Kalman smoothing and the expectation-maximisation algorithm}.
  \textit{Quarterly Journal of the Royal Meteorological Society},
  \textbf{143~(705)}, 1877--1885.

\bibitem[{Dun{\'{i}}k et~al.(2017)Dun{\'{i}}k, Straka, Kost,, and
  Havl{\'{i}}k}]{Dunik2017NoiseI}
Dun{\'{i}}k, J., O.~Straka, O.~Kost, and J.~Havl{\'{i}}k, 2017: {Noise
  covariance matrices in state-space models: A survey and comparison of
  estimation methods-Part I}. \textit{International Journal of Adaptive Control
  and Signal Processing}, \textbf{31~(11)}, 1505--1543.

\bibitem[{El~Gharamti(2018)}]{elgharamti2018enhanced}
El~Gharamti, M., 2018: {Enhanced Adaptive Inflation Algorithm for Ensemble
  Filters}. \textit{Monthly Weather Review}, \textbf{146}, 623--640.

\bibitem[{Evensen(2009)}]{evensen2009data}
Evensen, G., 2009: \textit{{Data assimilation: the ensemble Kalman filter}}.
  Springer Science {\&} Business Media.

\bibitem[{Frei and K{\"{u}}nsch(2012)Frei, and
  K{\"{u}}nsch}]{Frei2012SequentialFilters}
Frei, M., and H.~R. K{\"{u}}nsch, 2012: {Sequential State and Observation Noise
  Covariance Estimation Using Combined Ensemble Kalman and Particle Filters}.
  \textit{Monthly Weather Review}, \textbf{140~(5)}, 1476--1495.

\bibitem[{Fu et~al.(1993)Fu, Fukumori,, and Miller}]{Fu1993FittingModel}
Fu, L.-L., I.~Fukumori, and R.~N. Miller, 1993: {Fitting dynamic models to the
  Geosat sea level observations in the tropical Pacific ocean. Part II: A
  linear, wind-driven model}. \textit{Journal of Physical Oceanography},
  \textbf{23~(10)}, 2162--2181.

\bibitem[{Ghil and Malanotte-Rizzoli(1991)Ghil, and
  Malanotte-Rizzoli}]{Ghil1991DataOceanography.pdf}
Ghil, M., and P.~Malanotte-Rizzoli, 1991: {Data assimilation in meteorology and
  oceanography.pdf}. \textit{Advances in Geophysics}, \textbf{33}, 141--266.

\bibitem[{Guillet et~al.(2019)Guillet, Weaver, Vasseur, Michel, Gratton,, and
  G{\"{u}}rol}]{guillet2019modelling}
Guillet, O., A.~T. Weaver, X.~Vasseur, Y.~Michel, S.~Gratton, and
  S.~G{\"{u}}rol, 2019: {Modelling spatially correlated observation errors in
  variational data assimilation using a diffusion operator on an unstructured
  mesh}. \textit{Quarterly Journal of the Royal Meteorological Society},
  \doi{10.1002/qj.3537}.

\bibitem[{Harlim(2018)}]{Harlim2018EnsembleFilters}
Harlim, J., 2018: {Ensemble Kalman Filters}. \textit{Data-Driven Computational
  Methods}, Cambridge university press, 31--59.

\bibitem[{Harlim et~al.(2014)Harlim, Mahdi,, and Majda}]{Harlim2014AnModels}
Harlim, J., A.~Mahdi, and A.~J. Majda, 2014: {An ensemble Kalman filter for
  statistical estimation of physics constrained nonlinear regression models}.
  \textit{Journal of Computational Physics}, \textbf{257}, 782--812.

\bibitem[{Hollingsworth and L{\"{o}}nnberg(1986)Hollingsworth, and
  L{\"{o}}nnberg}]{hollingsworth1986statistical}
Hollingsworth, A., and P.~L{\"{o}}nnberg, 1986: {The statistical structure of
  short-range forecast errors as determined from radiosonde data. Part I: The
  wind field}. \textit{Tellus A}, \textbf{38~(2)}, 111--136.

\bibitem[{Hotta et~al.(2017)Hotta, Kalnay, Ota,, and Miyoshi}]{hotta2017}
Hotta, D., E.~Kalnay, Y.~Ota, and T.~Miyoshi, 2017: {EFSR: Ensemble forecast
  sensitivity to observation error covariance}. \textit{Monthly Weather
  Review}, \textbf{145}, 5015--5031.

\bibitem[{Houtekamer et~al.(2009)Houtekamer, Mitchell,, and
  Deng}]{Houtekamer2009ModelFilter}
Houtekamer, P.~L., H.~L. Mitchell, and X.~Deng, 2009: {Model Error
  Representation in an Operational Ensemble Kalman Filter}. \textit{Monthly
  Weather Review}, \textbf{137~(7)}, 2126--2143.

\bibitem[{Houtekamer and Zhang(2016)Houtekamer, and
  Zhang}]{Houtekamer2016ReviewAssimilation}
Houtekamer, P.~L., and F.~Zhang, 2016: {Review of the Ensemble Kalman Filter
  for Atmospheric Data Assimilation}. \textit{Monthly Weather Review},
  \textbf{144~(12)}, 4489--4532.

\bibitem[{Ide et~al.(1997)Ide, Courtier, Ghil,, and
  Lorenc}]{Ide1997UnifiedVariational}
Ide, K., P.~Courtier, M.~Ghil, and A.~C. Lorenc, 1997: {Unified Notation for
  Data Assimilation: Operational, Sequential and Variational}. \textit{Journal
  of the Meteorological Society of Japan}, \textbf{75~(1)}, 181--189.

\bibitem[{Janji{\'{c}} et~al.(2018)}]{Janjic2018OnAssimilation}
Janji{\'{c}}, T., and Coauthors, 2018: {On the representation error in data
  assimilation}. \textit{Quarterly Journal of the Royal Meteorological
  Society}, \textbf{144~(713)}, 1257--1278.

\bibitem[{Jazwinski(1970)}]{jazwinski2007stochastic}
Jazwinski, A.~H., 1970: \textit{{Stochastic processes and filtering theory}}.
  Academic Press.

\bibitem[{Kantas et~al.(2015)Kantas, Doucet, Singh, Maciejowski,, and
  Chopin}]{Kantas2015OnModels}
Kantas, N., A.~Doucet, S.~S. Singh, J.~Maciejowski, and N.~Chopin, 2015: {On
  particle methods for parameter estimation in state-space models}.
  \textit{Statistical Science}, \textbf{30~(3)}, 328--351.

\bibitem[{Katzfuss et~al.(2019)Katzfuss, Stroud,, and
  Wikle}]{katzfuss2019ensemble}
Katzfuss, M., J.~R. Stroud, and C.~K. Wikle, 2019: {Ensemble Kalman methods for
  high-dimensional hierarchical dynamic space-time models}. \textit{Journal of
  the American Statistical Association}, 1--43.

\bibitem[{Kotsuki et~al.(2017{\natexlab{a}})Kotsuki, Miyoshi, Terasaki, Lien,,
  and Kalnay}]{kotsuki2017assimilating}
Kotsuki, S., T.~Miyoshi, K.~Terasaki, G.-Y. Lien, and E.~Kalnay,
  2017{\natexlab{a}}: {Assimilating the global satellite mapping of
  precipitation data with the Nonhydrostatic Icosahedral Atmospheric Model
  (NICAM)}. \textit{Journal of Geophysical Research: Atmospheres},
  \textbf{122~(2)}, 631--650.

\bibitem[{Kotsuki et~al.(2017{\natexlab{b}})Kotsuki, Ota,, and
  Miyoshi}]{Kotsuki2017AdaptiveAtmosphere}
Kotsuki, S., Y.~Ota, and T.~Miyoshi, 2017{\natexlab{b}}: {Adaptive covariance
  relaxation methods for ensemble data assimilation: Experiments in the real
  atmosphere}. \textit{Quarterly Journal of the Royal Meteorological Society},
  \textbf{143~(705)}, 2001--2015.

\bibitem[{Leith(1978)}]{Leith1978ObjectivePrediction}
Leith, C.~E., 1978: {Objective methods for weather prediction}. \textit{Annual
  Review of Fluid Mechanics}, \textbf{10~(1)}, 107--128.

\bibitem[{Li et~al.(2009)Li, Kalnay,, and Miyoshi}]{Li2009SimultaneousFilter}
Li, H., E.~Kalnay, and T.~Miyoshi, 2009: {Simultaneous estimation of covariance
  inflation and observation errors within an ensemble Kalman filter}.
  \textit{Quarterly Journal of the Royal Meteorological Society},
  \textbf{135~(2)}, 523--533.

\bibitem[{Liang et~al.(2012)Liang, Zheng, Zhang, Wu, Dai,, and
  Li}]{Liang2012MaximumAssimilation}
Liang, X., X.~Zheng, S.~Zhang, G.~Wu, Y.~Dai, and Y.~Li, 2012: {Maximum
  likelihood estimation of inflation factors on error covariance matrices for
  ensemble Kalman filter assimilation}. \textit{Quarterly Journal of the Royal
  Meteorological Society}, \textbf{138~(662)}, 263--273.

\bibitem[{Liu and West(2001)Liu, and West}]{liu2001combined}
Liu, J., and M.~West, 2001: {Combined parameter and state estimation in
  simulation-based filtering}. \textit{Sequential Monte Carlo methods in
  practice}, Springer, 197--223.

\bibitem[{Liu et~al.(2017)Liu, Haussaire, Bocquet, Roustan, Saunier,, and
  Mathieu}]{Liu2017UncertaintyRadionuclides}
Liu, Y., J.-M. Haussaire, M.~Bocquet, Y.~Roustan, O.~Saunier, and A.~Mathieu,
  2017: {Uncertainty quantification of pollutant source retrieval: comparison
  of Bayesian methods with application to the Chernobyl and Fukushima Daiichi
  accidental releases of radionuclides}. \textit{Quarterly Journal of the Royal
  Meteorological Society}, \textbf{143~(708)}, 2886--2901.

\bibitem[{Mehra(1970)}]{Mehra1970OnFiltering}
Mehra, R.~K., 1970: {On the identification of variances and adaptive Kalman
  filtering}. \textit{IEEE Transactions on Automatic Control},
  \textbf{AC-15~(2)}, 175--184.

\bibitem[{Mehra(1972)}]{Mehra1972ApproachesFiltering}
Mehra, R.~K., 1972: {Approaches to adaptive filtering}. \textit{IEEE
  Transactions on Automatic Control}, \textbf{17~(5)}, 693--698.

\bibitem[{M{\'{e}}nard(2016)}]{Menard2016ErrorNetworks}
M{\'{e}}nard, R., 2016: {Error covariance estimation methods based on analysis
  residuals: theoretical foundation and convergence properties derived from
  simplified observation networks}. \textit{Quarterly Journal of the Royal
  Meteorological Society}, \textbf{142~(694)}, 257--273.

\bibitem[{Menemenlis and Chechelnitsky(2000)Menemenlis, and
  Chechelnitsky}]{Menemenlis2000ErrorData}
Menemenlis, D., and M.~Chechelnitsky, 2000: {Error estimates for an ocean
  general circulation model from altimeter and acoustic tomography data}.
  \textit{Monthly Weather Review}, \textbf{128~(3)}, 763--778.

\bibitem[{M{\'{e}}n{\'{e}}trier and Aulign{\'{e}}(2015)M{\'{e}}n{\'{e}}trier,
  and Aulign{\'{e}}}]{menetrier2015}
M{\'{e}}n{\'{e}}trier, B., and T.~Aulign{\'{e}}, 2015: {Optimized localization
  and hybridization to filter ensemble-based covariances}. \textit{Monthly
  Weather Review}, \textbf{143}, 3931--3947.

\bibitem[{Merchant et~al.(2020)Merchant, Saux-Picart,, and
  Waller}]{Merchant2020BiasReferences}
Merchant, C.~J., S.~Saux-Picart, and J.~Waller, 2020: {Bias correction and
  covariance parameters for optimal estimation by exploiting matched in-situ
  references}. \textit{Remote Sensing of Environment}, \textbf{237}, 111\,590.

\bibitem[{Mitchell and Houtekamer(2000)Mitchell, and
  Houtekamer}]{Mitchell2000AnFilter}
Mitchell, H.~L., and P.~L. Houtekamer, 2000: {An adaptive ensemble Kalman
  filter}. \textit{Monthly Weather Review}, \textbf{128~(2)}, 416--433.

\bibitem[{Mitchell and Carrassi(2015)Mitchell, and
  Carrassi}]{Mitchell2015AccountingFiltering}
Mitchell, L., and A.~Carrassi, 2015: {Accounting for model error due to
  unresolved scales within ensemble Kalman filtering}. \textit{Quarterly
  Journal of the Royal Meteorological Society}, \textbf{141~(689)}, 1417--1428.

\bibitem[{Miyoshi(2011)}]{Miyoshi2011TheFilter}
Miyoshi, T., 2011: {The Gaussian Approach to Adaptive Covariance Inflation and
  Its Implementation with the Local Ensemble Transform Kalman Filter}.
  \textit{Monthly Weather Review}, \textbf{139~(5)}, 1519--1535.

\bibitem[{Miyoshi et~al.(2013)Miyoshi, Kalnay,, and
  Li}]{Miyoshi2013EstimatingAssimilation}
Miyoshi, T., E.~Kalnay, and H.~Li, 2013: {Estimating and including
  observation-error correlations in data assimilation}. \textit{Inverse
  Problems in Science and Engineering}, \textbf{21~(3)}, 387--398.

\bibitem[{Miyoshi et~al.(2010)Miyoshi, Sato,, and
  Kadowaki}]{Miyoshi2010EnsembleSystem}
Miyoshi, T., Y.~Sato, and T.~Kadowaki, 2010: {Ensemble Kalman filter and 4D-Var
  intercomparison with the Japanese operational global analysis and prediction
  system}. \textit{Monthly Weather Review}, \textbf{138~(7)}, 2846--2866.

\bibitem[{Pham et~al.(1998)Pham, Verron,, and Roubaud}]{pham1998singular}
Pham, D.~T., J.~Verron, and M.~C. Roubaud, 1998: {A singular evolutive extended
  Kalman filter for data assimilation in oceanography}. \textit{Journal of
  Marine systems}, \textbf{16~(3-4)}, 323--340.

\bibitem[{Pulido et~al.(2018)Pulido, Tandeo, Bocquet, Carrassi,, and
  Lucini}]{Pulido2018StochasticMethods}
Pulido, M., P.~Tandeo, M.~Bocquet, A.~Carrassi, and M.~Lucini, 2018:
  {Stochastic parameterization identification using ensemble Kalman filtering
  combined with maximum likelihood methods}. \textit{Tellus A: Dynamic
  Meteorology and Oceanography}, \textbf{70~(1)}, 1442\,099.

\bibitem[{Purser and Parrish(2003)Purser, and
  Parrish}]{Purser2003AAssimilation}
Purser, R.~J., and D.~F. Parrish, 2003: {A Bayesian technique for estimating
  continuously varying statistical parameters of a variational assimilation}.
  \textit{Meteorology and Atmospheric Physics}, \textbf{82~(1-4)}, 209--226.

\bibitem[{Raanes et~al.(2019)Raanes, Bocquet,, and
  Carrassi}]{raanes2019apdative}
Raanes, P.~N., M.~Bocquet, and A.~Carrassi, 2019: {Adaptive covariance
  inflation in the ensemble Kalman filter by Gaussian scale mixtures}.
  \textit{Quarterly Journal of the Royal Meteorological Society},
  \textbf{145~(718)}, 53--75.

\bibitem[{Rodwell and Palmer(2007)Rodwell, and Palmer}]{rodwell2007using}
Rodwell, M.~J., and T.~N. Palmer, 2007: {Using numerical weather prediction to
  assess climate models}. \textit{Quarterly Journal of the Royal Meteorological
  Society}, \textbf{133~(622)}, 129--146.

\bibitem[{Ruiz et~al.(2013{\natexlab{a}})Ruiz, Pulido,, and
  Miyoshi}]{Ruiz2013EstimatingReview}
Ruiz, J.~J., M.~Pulido, and T.~Miyoshi, 2013{\natexlab{a}}: {Estimating model
  parameters with ensemble-based data assimilation: A review}. \textit{Journal
  of the Meteorological Society of Japan}, \textbf{91}, 79--99.

\bibitem[{Ruiz et~al.(2013{\natexlab{b}})Ruiz, Pulido,, and
  Miyoshi}]{Ruiz2013EstimatingTreatment}
Ruiz, J.~J., M.~Pulido, and T.~Miyoshi, 2013{\natexlab{b}}: {Estimating model
  parameters with ensemble-based data assimilation: Parameter Covariance
  Treatment}. \textit{Journal of the Meteorological Society of Japan},
  \textbf{91}, 453--469.

\bibitem[{Rutherford(1972)}]{rutherford1972data}
Rutherford, I.~D., 1972: {Data assimilation by statistical interpolation of
  forecast error fields}. \textit{Journal of the Atmospheric Sciences},
  \textbf{29~(5)}, 809--815.

\bibitem[{Satterfield et~al.(2018)Satterfield, Hodyss, Kuhl,, and
  Bishop}]{satterfield2018}
Satterfield, E.~A., D.~Hodyss, D.~D. Kuhl, and C.~H. Bishop, 2018:
  {Observation-informed generalized hybrid error covariance models}.
  \textit{Monthly Weather Review}, \textbf{146}, 3605--3622.

\bibitem[{Scheffler et~al.(2019)Scheffler, Ruiz,, and
  Pulido}]{scheffler2018inference}
Scheffler, G., J.~Ruiz, and M.~Pulido, 2019: {Inference of stochastic
  parametrizations for model error treatment using nested ensemble Kalman
  filters}. \textit{Quarterly Journal of the Royal Meteorological Society},
  \doi{https://doi.org/10.1002/qj.3542}.

\bibitem[{Schmidt(1966)}]{schmidt1966applications}
Schmidt, S.~F., 1966: {Applications of state space methods to navigation
  problems}. \textit{Advances in Control Systems}, \textbf{3}, 293--340.

\bibitem[{Shumway and Stoffer(1982)Shumway, and
  Stoffer}]{Shumway1982AnAlgorithm}
Shumway, R.~H., and D.~S. Stoffer, 1982: {An approach to time series smoothing
  and forecasting using the EM algorithm}. \textit{Journal of Time Series
  Analysis}, \textbf{3~(4)}, 253--264.

\bibitem[{Solonen et~al.(2014)Solonen, Hakkarainen, Ilin, Abbas,, and
  Bibov}]{Solonen2014EstimatingFiltering}
Solonen, A., J.~Hakkarainen, A.~Ilin, M.~Abbas, and A.~Bibov, 2014: {Estimating
  model error covariance matrix parameters in extended Kalman filtering}.
  \textit{Nonlinear Processes in Geophysics}, \textbf{21~(5)}, 919--927.

\bibitem[{Stroud and Bengtsson(2007)Stroud, and
  Bengtsson}]{Stroud2007SequentialFilter}
Stroud, J.~R., and T.~Bengtsson, 2007: {Sequential state and variance
  estimation within the ensemble Kalman filter}. \textit{Monthly Weather
  Review}, \textbf{135~(9)}, 3194--3208.

\bibitem[{Stroud et~al.(2018)Stroud, Katzfuss,, and
  Wikle}]{Stroud2018AEstimation}
Stroud, J.~R., M.~Katzfuss, and C.~K. Wikle, 2018: {A Bayesian adaptive
  ensemble Kalman filter for sequential state and parameter estimation}.
  \textit{Monthly Weather Review}, \textbf{146~(1)}, 373--386.

\bibitem[{Tandeo et~al.(2015)Tandeo, Pulido,, and Lott}]{tandeo2015offline}
Tandeo, P., M.~Pulido, and F.~Lott, 2015: {Offline parameter estimation using
  EnKF and maximum likelihood error covariance estimates: Application to a
  subgrid-scale orography parametrization}. \textit{Quarterly Journal of the
  Royal Meteorological Society}, \textbf{141~(687)}, 383--395.

\bibitem[{Todling(2015)}]{Todling2015AMethod}
Todling, R., 2015: {A lag-1 smoother approach to system-error estimation:
  sequential method}. \textit{Quarterly Journal of the Royal Meteorological
  Society}, \textbf{141~(690)}, 1502--1513.

\bibitem[{Ueno et~al.(2010)Ueno, Higuchi, Kagimoto,, and
  Hirose}]{Ueno2010MaximumModel}
Ueno, G., T.~Higuchi, T.~Kagimoto, and N.~Hirose, 2010: {Maximum likelihood
  estimation of error covariances in ensemble-based filters and its application
  to a coupled atmosphere-ocean model}. \textit{Quarterly Journal of the Royal
  Meteorological Society}, \textbf{136~(650)}, 1316--1343.

\bibitem[{Ueno and Nakamura(2014)Ueno, and Nakamura}]{Ueno2014IterativeFilters}
Ueno, G., and N.~Nakamura, 2014: {Iterative algorithm for maximum-likelihood
  estimation of the observation-error covariance matrix for ensemble-based
  filters}. \textit{Quarterly Journal of the Royal Meteorological Society},
  \textbf{140~(678)}, 295--315.

\bibitem[{Ueno and Nakamura(2016)Ueno, and Nakamura}]{Ueno2016BayesianFilters}
Ueno, G., and N.~Nakamura, 2016: {Bayesian estimation of the observation-error
  covariance matrix in ensemble-based filters}. \textit{Quarterly Journal of
  the Royal Meteorological Society}, \textbf{142~(698)}, 2055--2080.

\bibitem[{Wahba and Wendelberger(1980)Wahba, and Wendelberger}]{wahba1980some}
Wahba, G., and J.~Wendelberger, 1980: {Some new mathematical methods for
  variational objective analysis using splines and cross validation}.
  \textit{Monthly weather review}, \textbf{108~(8)}, 1122--1143.

\bibitem[{Waller et~al.(2016)Waller, Dance,, and
  Nichols}]{Waller2016TheoreticalStatistics}
Waller, J.~A., S.~L. Dance, and N.~K. Nichols, 2016: {Theoretical insight into
  diagnosing observation error correlations using observation-minus-background
  and observation-minus-analysis statistics}. \textit{Quarterly Journal of the
  Royal Meteorological Society}, \textbf{142~(694)}, 418--431.

\bibitem[{Waller et~al.(2017)Waller, Dance,, and
  Nichols}]{waller2017diagnosing}
Waller, J.~A., S.~L. Dance, and N.~K. Nichols, 2017: {On diagnosing
  observation-error statistics with local ensemble data assimilation}.
  \textit{Quarterly Journal of the Royal Meteorological Society},
  \textbf{143~(708)}, 2677--2686.

\bibitem[{Wang and Bishop(2003)Wang, and Bishop}]{Wang2003ASchemes}
Wang, X., and C.~H. Bishop, 2003: {A Comparison of Breeding and Ensemble
  Transform Kalman Filter Ensemble Forecast Schemes}. \textit{Journal of the
  Atmospheric Sciences}, \textbf{60~(9)}, 1140--1158.

\bibitem[{Weston et~al.(2014)Weston, Bell,, and Eyre}]{weston2014accounting}
Weston, P.~P., W.~Bell, and J.~R. Eyre, 2014: {Accounting for correlated error
  in the assimilation of high-resolution sounder data}. \textit{Quarterly
  Journal of the Royal Meteorological Society}, \textbf{140~(685)}, 2420--2429.

\bibitem[{Whitaker and Hamill(2012)Whitaker, and
  Hamill}]{Whitaker2012EvaluatingAssimilation}
Whitaker, J.~S., and T.~M. Hamill, 2012: {Evaluating methods to account for
  system errors in ensemble data assimilation}. \textit{Monthly Weather
  Review}, \textbf{140~(9)}, 3078--3089.

\bibitem[{Whitaker et~al.(2008)Whitaker, Hamill, Wei, Song,, and
  Toth}]{Whitaker2008EnsembleSystem}
Whitaker, J.~S., T.~M. Hamill, X.~Wei, Y.~Song, and Z.~Toth, 2008: {Ensemble
  data assimilation with the NCEP global forecast system}. \textit{Monthly
  Weather Review}, \textbf{136~(2)}, 463--482.

\bibitem[{Winiarek et~al.(2014)Winiarek, Bocquet, Duhanyan, Roustan, Saunier,,
  and Mathieu}]{Winiarek2014EstimationObservations}
Winiarek, V., M.~Bocquet, N.~Duhanyan, Y.~Roustan, O.~Saunier, and A.~Mathieu,
  2014: {Estimation of the caesium-137 source term from the Fukushima Daiichi
  nuclear power plant using a consistent joint assimilation of air
  concentration and deposition observations}. \textit{Atmospheric Environment},
  \textbf{82}, 268--279.

\bibitem[{Winiarek et~al.(2012)Winiarek, Bocquet, Saunier,, and
  Mathieu}]{Winiarek2012EstimationPlant}
Winiarek, V., M.~Bocquet, O.~Saunier, and A.~Mathieu, 2012: {Estimation of
  errors in the inverse modeling of accidental release of atmospheric
  pollutant: Application to the reconstruction of the cesium-137 and iodine-131
  source terms from the Fukushima Daiichi power plant}. \textit{Journal of
  Geophysical Research: Atmospheres}, \textbf{117~(D5)}.

\bibitem[{Wu(1983)}]{Wu1983OnAlgorithm}
Wu, C. F.~J., 1983: {On the convergence properties of the EM algorithm}.
  \textit{Annals of Statistics}, \textbf{11~(1)}, 95--103.

\bibitem[{Yang and M{\'{e}}min(2019)Yang, and M{\'{e}}min}]{yang2019estimation}
Yang, Y., and E.~M{\'{e}}min, 2019: {Estimation of physical parameters under
  location uncertainty using an ensemble-expectation-maximization algorithm}.
  \textit{Quarterly Journal of the Royal Meteorological Society},
  \textbf{145~(719)}, 418--433.

\bibitem[{Ying and Zhang(2015)Ying, and Zhang}]{Ying2015AnAssimilation}
Ying, Y., and F.~Zhang, 2015: {An adaptive covariance relaxation method for
  ensemble data assimilation}. \textit{Quarterly Journal of the Royal
  Meteorological Society}, \textbf{141~(692)}, 2898--2906.

\bibitem[{Zebiak and Cane(1987)Zebiak, and Cane}]{zebiak1987model}
Zebiak, S.~E., and M.~A. Cane, 1987: {A model El Nino--southern oscillation}.
  \textit{Monthly Weather Review}, \textbf{115~(10)}, 2262--2278.

\bibitem[{Zhang et~al.(2004)Zhang, Snyder,, and Sun}]{Zhang2004ImpactsFilter}
Zhang, F., C.~Snyder, and J.~Sun, 2004: {Impacts of initial estimate and
  observation availability on convective-scale data assimilation with an
  ensemble Kalman filter}. \textit{Monthly Weather Review}, \textbf{132~(5)},
  1238--1253.

\bibitem[{Zhen and Harlim(2015)Zhen, and Harlim}]{Zhen2015AdaptiveFilters}
Zhen, Y., and J.~Harlim, 2015: {Adaptive error covariances estimation methods
  for ensemble Kalman filters}. \textit{Journal of Computational Physics},
  \textbf{294}, 619--638.

\end{thebibliography}

\end{document}